\documentclass  [preprint,showpacs,preprintnumbers,amsmath,amssymb]{revtex4}

\usepackage{graphicx}
\usepackage{dcolumn}
\usepackage{bm}
\usepackage{epsfig}

\begin{document}
\newcommand{\gsim}{\hbox{\rlap{$^>$}$_\sim$}}
\newcommand{\lsim}{\hbox{\rlap{$^<$}$_\sim$}}

\title{Critical Tests Of The Leading Gamma Ray Burst Theories}

\author{Shlomo Dado and Arnon Dar}
\affiliation{Physics Department, Technion, Haifa 32000, Israel}

\begin{abstract}
Although it has been established observationally beyond doubt that broadline 
stripped envelope supernovae (SNe) of type Ic produce long duration gamma ray 
bursts (GRBs), that neutron star mergers produce short hard GRBs (SHBs), and that 
phase transition of neutron stars in high mass X-ray binaries (HMXBs) may produce 
SN-Less GRBs, their production mechanism is still debated. The two leading 
theoretical models of GRBs and their afterglows, the fireball model and the 
cannonball model, have been widely confronted with the mounting observational 
data on GRBs and SHBs during the last two decades. Both have claimed success in 
reproducing the observational data, despite their complexity and diversity. This 
claimed success, however, may reflect multiple choices and the use of many free 
adjustable parameters, rather than the true validity of the models. Only 
confrontation of the key falsifiable predictions of the models with solid 
observational data can test their validity. Such critical tests are reviewed in 
this report.
\end{abstract}


\maketitle

\section{Introduction} 

Gamma-ray bursts (GRBs) are brief flashes of gamma rays lasting between 
few milliseconds and several hours from extremely energetic cosmic 
explosions [1]. They were first detected on July 2, 1967 by the USA Vela 
spy satellites, which were launched to detect USSR tests of 
nuclear weapons above the atmosphere, in violation of the USA-USSR 
Nuclear Test Ban Treaty signed in 1963. Their discovery was first 
published in 1973 after 15 such events were detected [2], which 
have ruled out man-made origin and indicated that they were outside 
the solar system. 

During the first 20 years after their discovery, hundreds of models of 
GRBs were published (see, e.g., [3]), where it was assumed that GRBs are 
Galactic in origin. An extragalactic origin implied implausible energy 
release in gamma rays from a very small volume in a very short time, if 
they were isotropic, as was generally assumed. During that period it was 
also found that GRBs fall roughly into two classes, long duration ones 
(LGRBs) that last more than $\sim$2 seconds, and short bursts (SGRBs) 
that typically last less than $\sim$2 seconds [4] most of which are 
short hard bursts (SHBs) with a spectrum much harder than LGRBs. The 
origin and production mechanism of both types of GRBs have been major 
astrophysical puzzles until recently.

In 1984, Blinnikov et al. [5] suggested that exploding neutron stars in 
close binaries may produce GRBs with isotropic gamma ray energy that 
could reach $\sim\! 10^{46}$ erg. Such GRBs could be seen only from 
relatively nearby galaxies. Paczynsky, however, maintained [6] that the 
sky distribution of GRBs is more consistent with large cosmological 
distances, like those of quasars, with a redshift of about 1 or 2, which 
implies a release of supernova-like energy, $\sim\! 10^{51}$ erg, within 
less than 1 s, making gamma-ray bursters the brightest objects known in 
the universe, many orders of magnitude brighter than any quasar [6].

The first plausible physical model of extragalactic gamma-ray bursts at 
large cosmological distances, was proposed by Goodman, Dar and Nussinov 
in 1987 [7]. They suggested that extragalactic GRBs may 
be produced in stripped envelope supernova explosions (SNe) and in 
neutron stars mergers  by an $e^+e^-\gamma$ fireball [8] formed by 
neutrino-antineutrino annihilation around the newly born compact object 
-- a massive neutron star (n*) or a black hole (bh). But shortly after 
the launch of the Compton Gamma-Ray Burst Observatory (CGRO) in 1991, it 
became clear that such neutrino-annihilation fireballs are not powerful 
enough to produce observable GRBs at the very large cosmological 
distances, which were indicated by the CGRO observations [9], unless the 
produced $e^+e^-\gamma$ fireballs are collimated by funneling through 
surrounding matter into a conical fireball [10]. Shaviv and Dar, 
however, suggested instead [11] that narrowly beamed GRBs can be 
produced by jets of highly relativistic plasmoids (cannonballs) of 
ordinary matter through inverse Compton scattering (ICS) of light 
surrounding their launch sites. They proposed that such jets may be 
launched in stripped-envelope core-collapse supernova explosions, in 
merger of compact stars due to the emission of gravitational waves, and 
in phase transition of neutron stars to a more compact object, i.e., a 
quark star (q*) or a black hole (bh), following mass accretion in 
compact binaries. 

An important prediction of the fireball model was a transition of the initial 
short $\gamma$-ray emission to emission of a longer-lived "afterglow" [12] at 
longer wavelengths due to the slow down of the expansion of the 
$e^+e^-\,\gamma$ fireball by the swept in surrounding medium. In 1997, 
measurements with the Italian-Dutch satellite BeppoSAX discovered that GRBs are 
indeed followed by a longer-lived X-ray "afterglow" [13]. It provided accurate 
enough sky localization of GRBs, which led to the discovery of their afterglow 
at longer wavelengths [14], the discovery of their host galaxies [15] and their 
redshifts [16] shortly after, and the association of long GRBs with supernova 
(SN) explosions of type Ic [17]. Following measurements during the past 20 
years, mainly with the X-ray satellites HETE, Swift, Konus-Wind, Chandra, 
Integral, XMM-Newton, and Fermi, the Hubble space telescope, and ground based 
telescopes, provided the detailed properties of the prompt and afterglow 
emissions of GRBs over the entire electromagnetic spectrum, the association 
of  LGRBs with SNeIc and the properties of their host galaxies and near 
environments [18]. In particular, they provided clear evidence that LGRBs  
are taking place mainly in star formation regions within the disk of
spiral galaxies, where most SNeIc take place, while SGRBs are taking place 
in and outside of both spiral and  elliptical galaxies. This suggested a 
different origin of LGRBs and SGRBs. While LGRBs were observed to be 
associated with SNeIc, SHBs were not. That, and the location of SHBs  
led to the wide spread belief that SHBs are produced in merger of neutron 
stars, and/or a  neutron star and a black hole [7,10] in close binaries.

This belief was based on indirect evidence [19]. Recently, however, SHB170817A 
[20] that followed $1.74\!\pm\! 0.05$ s after the gravitation wave (GW) chirp 
from a relatively nearby neutron stars merger (NSM) event, GW170817, which was 
detected with the Ligo-Virgo GW detectors [21], has shown beyond doubt that 
neutron star mergers produce SHBs. Moreover, the universal shape of all the 
well sampled early time afterglow of ordinary SHBs and of SHB170817A [22], 
which is expected from a pulsar wind nebula (PWN) emission powered by the spin 
down of a newly born milli second pulsar (MSP), suggest that most SHBs, are 
produced by NSMs yielding a neutron star remnant rather than a stellar mass 
black hole [22].

Although long duration nearby GRBs have been seen in association with 
very bright broad-line supernova explosions of type Ic [17,23], no 
associated SN has been detected in several nearby long duration 
GRBs despite very deep searches [24]. The universal behavior of the 
afterglow of such long duration SN-Less GRBs and SHBs 
[22, 25], however, suggest that they are also powered by a newly born 
millisecond pulsars, perhaps in phase transition of neutron stars 
to quark stars [11,26] following mass accretion onto neutron stars in 
high mass X-ray binaries (HMXBs).

Since 1997 only two theoretical models of GRBs and their 
afterglows, the standard fireball (FB) model [27] and the 
cannonball (CB) model [28], have been used extensively to 
interpret the mounting observational data on GRBs and their 
afterglows. Both models have claimed to reproduce well the 
observational data. But, despite their similar names, the two 
models were originally and still are very different in their basic 
assumptions and predictions. This is despite the  
replacement of key assumptions underlying the standard FB models with 
assumptions underlying the CB model (see below). The claimed 
success, however, of both  models  
in reproducing the mounting observational data on GRBs and 
their afterglows, despite the complexity and diversity of these 
data, may reflect the fact that the predictions of both models 
depend on free parameters and choices, which, for each GRB, 
were adjusted to fit the observational data. As a result, when 
successful fits to observational data were obtained, it was not 
clear whether they were due to the validity of the theory or due to 
multiple choices and the use of many adjustable parameters to 
describe individual GRBs and their afterglows.

Scientific theories, however, must be falsifiable [29]. Hence, only 
confrontations between key predictions of the GRB models, which do 
not depend on free adjustable parameters, with solid observational 
data can serve as critical tests of the validity of such models, 
rather than biases, prejudices, consensus or beliefs. Such critical tests of 
the cannonball model and the standard fireball model of long GRBs 
and SHBs are reviewed in this report. The obvious conclusion is left 
to be drawn by the unbiased reader.  

\section{The GRB Models} 
GRBs and SHBs seem to consist of a few short $\gamma$-ray pulses 
with roughly a fast rise and an exponential decay (FRED) pulse-shape [1]. 
The number of such individual pulses, their 
time sequence, relative intensities, and durations, that vary 
drastically from burst to burst and within bursts, are not
predicted by the GRB models. The main properties of resolved pulses, however, 
such as pulse-shape, polarization and correlations between their main 
properties, as well as global properties of the entire bursts, 
and their afterglows are predicted by the models and can be used for 
critical tests of the modes. Since LGRBs and SGRBs have different progenitors
we shall discuss the critical tests of their CB and FB models separately.
\begin{figure}[]
\centering
\epsfig{file=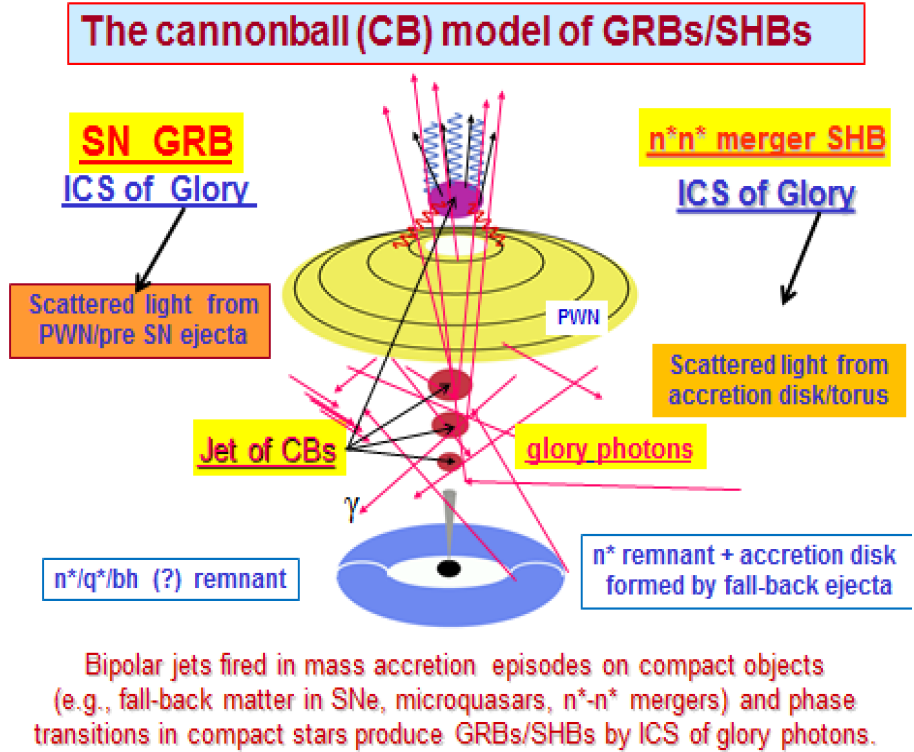,width=12.cm,height=10.cm}
\caption{Schematic description of the CB model of GRBs}
\end{figure}
{\bf The CB model} [28,30,26,25], is illustrated in Figure 1.
In the CB model,  bipolar jets of highly relativistic plasmoids 
(cannonballs) are assumed to be launched by fall back material onto the 
newly born compact stellar object [11], a neutron star, a quark 
star or a black hole in stripped envelope supernovae explosions 
of type Ic (SN-GRBs), in NSM in close binaries (SHBs), and in phase 
transition of neutron stars to q* due to mass accretion (SN-less GRBs) in 
HMXBs. The prompt emission $\gamma$-ray pulses 
are produced by ICS of the radiation (glory) 
surrounding the launch site, by the electrons enclosed in the CBs 
of the jet. In SN-GRBs, this glory can be the light halo formed around the 
progenitor star by scattered light from pre-supernova ejections. 
In SN-less GRBs it can be light from the massive star companion, 
or the radiation emitted from the accretion disk formed around the 
neutron star. In SHBs it can be the X-ray radiation  
from an accretion disk formed around the n*s remnant by fall back of 
tidally disrupted material or  debris from the final  explosion of 
the lighter n* [5] after loosing most of its mass.  

When the CBs enter the interstellar medium, they decelerate by 
sweeping in the ionized medium in front of them. The swept in 
electrons and nuclei are Fermi accelerated there to very high 
energies by the turbulent magnetic fields present/generated in the 
CBs. The accelerated electrons cool mainly by emitting synchrotron 
radiation, which dominates the afterglow of SN-GRBs that usually 
take place in dense stellar regions - molecular clouds where most 
SNe take place. 

The afterglows of  SN-less GRBs and SHBs, which are 
usually produced in much lower density environments than those of
SN-GRBs, appear to be  dominated by the radiation from a pulsar wind 
nebula, which is powered by the spin down of the newly born 
millisecond pulsar [25,26].\\

\begin{figure}[]
\centering
\epsfig{file=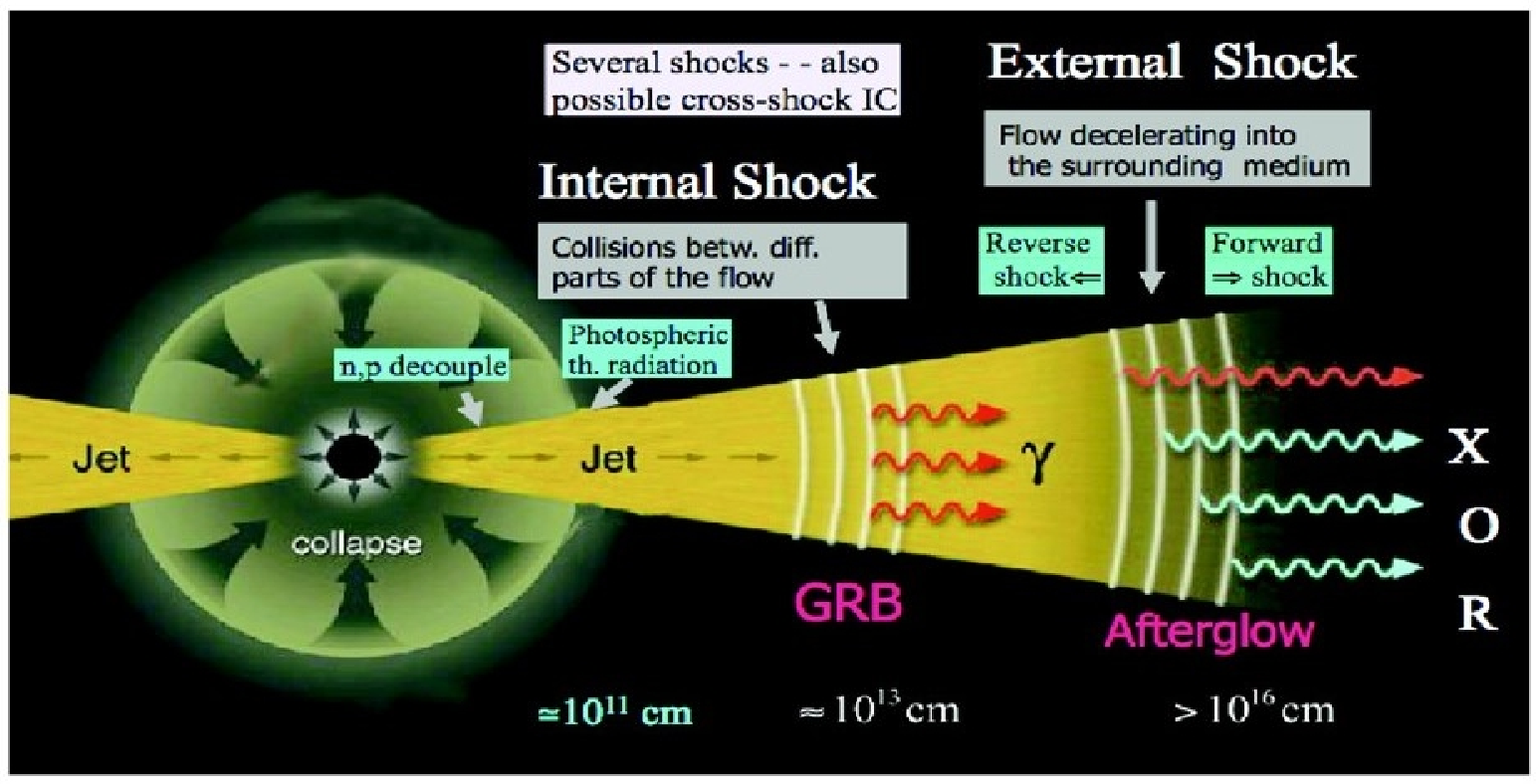,width=15cm,height=8.cm}
\caption{Schematic description of the fireball model of GRBs 
adapted from  a recent review of GRBs by Meszaros and Rees [31].}
\end{figure}

{\bf The FB models of GRBs} evolved a long way from the original 
spherical $e^+e^-\gamma$ fireballs [8] formed around  stripped 
envelope supernova explosions [7], n*n* mergers [7]
and n*bh mergers [10] to the current collimated fireball models [27]. 
The  most popular version is illustrated in Figure 2 adapted from  [31].
It assumes that long GRBs are produced by highly relativistic 
conical jets of ordinary matter launched by collapsars - the collapse 
of a massive star to a black hole,  either directly without a supernova 
("failed supernova") [32, 33], or indirectly in  a hypernova (the delayed 
collapse  of the newly born compact object  to a black hole by mass accretion 
of fall back material in core collapse supernova explosion[34]). 
SHBs are assumed to be produced by 
highly relativistic jets launched in n*n* and n*bh mergers.

In  the FB models, the prompt emission pulses are assumed to be produced 
by synchrotron radiation emitted by highly relativistic electrons 
shock accelerated in the collisions between overtaking conical shells. 
The continuous collision of the merged shells with the 
circumburst medium is assumed to drive a forward shock into the 
interstellar  medium or pre-ejected stellar wind, 
and a reverse shock into the merged shells. 
The shock accelerated  electrons produce synchrotron 
radiation (SR) afterglow [27] on top of a hypernova [34] light in 
LGRBs, or a macronova light in SHBs [35]. The reverse shock produces 
the optical photons while inverse Compton scattering of the SR in the 
forward blast wave produce GeV-TeV photons.  

\section{Prompt emission tests}
\noindent{\bf Test 1: Polarization.}\\
{\bf In the CB model}, ordinary GRBs  are 
produced by narrowly collimated  jets of CBs with a bulk motion Lorentz 
factor $\gamma_0\!=\!\gamma(t\!=\!0)\!\gg\!1$ through inverse Compton
scattering of light. They  are narrowly beamed and are viewed 
from  small angles $\theta\!\approx\! 1/\gamma_0\!\ll\!1$ relative 
to the jet direction of motion, i.e., with  Doppler factors
$\delta_0\!=\!\delta(t\!=\!0)\!=\!1/\gamma_0(1\!-\!\beta\,cos\theta)\!
\approx\! 2\gamma_0/(1\!+\!\gamma_0^2\theta^2)$ to a good approximation
for $\gamma_0^2\!\gg\!1$, and $\theta^2\!\ll\!1$. 
For the most probable viewing angles $\theta\!\approx 1/\gamma_0$
of such GRBs, their expected linear polarization is [36]:
\begin{equation}
\Pi\!=\!2\gamma_0^2\,\theta^2/(1\!+\!\gamma_0^4\,\theta^4)\!\approx\! 100\%.
\end{equation}
High luminosity (HL) GRBs, or low luminosity (LL) GRBs, that in the 
CB model are mainly GRBs viewed from very near axis 
($\gamma_0^2\,\theta^2\!\ll\!1$), or very far off-axis 
($\gamma_0^2\,\theta^2\!\gg\!1$), respectively, are expected to display 
a small linear polarization. For instance, $\Pi\!<\! 0.22$ is predicted 
for HL GRBs with $\gamma_0\,\theta\!<\!1/3$, and for LL GRBs with 
$\gamma_0\,\theta\!>\!3$. However, HL or LL GRBs that are very bright 
or very dim, respectively, because of having an unusual very large or 
very small $\gamma_0$, respectively, and are viewed from 
$\theta\!\approx\!1/\gamma_0$, are expected to display a rather large 
polarization.

\noindent 
{\bf In the standard FB models}, GRB pulses are produced by synchrotron 
radiation emitted by high energy electrons, which are Fermi/shock 
accelerated in collisions between conical shells or by shocks within 
conical flows (jets). Such Fermi/shock acceleration, however, requires 
highly turbulent magnetic fields in the acceleration region, which 
produce a very small net polarization.  Indeed, the afterglow of GRBs, 
that  in both  the CB model and the FB models is produced by 
synchrotron emission from shock accelerated electrons, is observed 
to have a  very small polarization [37]. This is in contrast to the 
linear polarization of the prompt emission, which has been 
found to be very large in all GRBs where it was measured [38], as summarized 
in Table I. Soon after the first report of a measurement of a large 
polarization of the prompt emission in a GRB021206, observers 
questioned the measurement while promoters of firecone/fireshell models 
proposed posteriori explanations. For instance, it was suggested that 
a constant magnetic field exists in the small domains of an angular 
size $\sim\! 1/\gamma_0$ of the firecone/fireshell from where the photons 
arrive simultaneously in the observer frame. However, 
the observed photons  at any given time are a collection of photons,
which have the same arrival time, but not the same emission time, i.e.,  
the magnetic field in these different emission domains  
must align in nearly in the same direction. Such a situation 
must be present in most GRBs in order     
to explain the large observed polarization in all cases where it was 
measured [38]. However, a highly turbulent magnetic field rather 
than an ordered field is needed in order to Fermi/shock accelerate 
the electrons to high energy whose emitted synchrotron radiation is assumed to 
be the GRB prompt emission. Moreover, the above explanations of the observed 
large polarization [38] are  in tension with the curvature effect [39] 
which was claimed  to explain the observed temporal and spectral behaviors 
of the prompt emission pulses, and with the relatively small polarization of 
the afterglow observed right after the prompt emission [37] - an hypothesized  
constant magnetic field within domains of a size $\approx \!1/\gamma$ would 
produce also a large polarization during the afterglow phase in contradiction 
to the small observed  polarization. 
\begin{table*}
\caption{GRBs with measured polarization of prompt $\gamma$-rays}
\label{table1}
\centering
\begin{tabular}{l l l l l l}
\hline
\hline
~~~GRB~~~ & Polarization(\%) & ~CL~ &~~~Reference~[38] ~~~ & Polarimetry~~  \\
~021206~~ & 80\,+/-\,20      & ???  & Coburn \& Boggs 2003 & RHESSI~~~~~~~  \\
~930131~~ & \,\,\,$>\!35$,   & 90\% & Willis et al.   2005 & BATSE (Albedo) \\
~960924~~ & \,\,\,$>\!50$    & 90\% & Willis et al.   2005 & BATSE (ALbedo) \\
041219A~~ & \,\,98\,+/-\,33  & 68\% & Kalemci et al.  2007 & INTEGRAL-SPI~~ \\
100826A~~ & 27\,+/-\,11      & 99\% & Yonetoku et al. 2011 & IKARUS-GAP~~~~ \\
110301A~~ & 70\,+/-\,22      & 68\% & Yonetoku et al. 2012 & IKARUS-GAP~~~~ \\
~110721~~ & 84\,+16/-28      & 68\% & Yonetoku et al. 2012 & IKARUS-GAP~~~~ \\
~061122~~ & \,\,\, $>\!60$   & 68\% & Gotz et al.     2013 & INTEGRAL-IBIS~ \\
140206A   & \,\,\, $>\!48$   & 68\% & Gotz et al. ~~~~2014 & INTEGRAL-IBIS~ \\
\hline
\end{tabular}
\end{table*}

\noindent
{\bf Test 2: Correlations.}\\
{\bf The CB model} entails very simple correlations between the main
observables of GRBs [40]. For instance, ICS of 
glory photons of energy $\epsilon$ by CBs 
boosts their energy to $E_\gamma\!=\!\gamma_0\,\delta_0\epsilon/(1+z)$
in the observer frame. Consequently, the peak energy $E_p$ 
of their  time-integrated energy distribution  satisfies,
\begin{equation}
(1+z)\,E_p\!\propto\! \gamma_0\,\delta_0\, \epsilon_p\,, 
\end{equation}
where $\epsilon_p$ is the peak energy of the glory.
In the Thomson regime, the nearly isotropic distribution of the ICS 
photons in the CB rest frame (primed) is beamed into a distribution
$dn_\gamma/d\omega\!=\!(dn'_\gamma/d\omega)\, \delta^2\!\approx\! 
(n_\gamma/4\,\pi)\,\delta^2$ in the observer frame.
Consequently, the isotropic-equivalent total  energy of such scattered 
photons satisfies 
\begin{equation}
E_{iso}\!\propto\! \gamma_0\, \delta_0^3\, \epsilon_p. 
\end{equation}
Hence, both ordinary LGRBs and SGRBs, 
which in the CB model are GRBs viewed mostly from an angle $\theta\approx 
1/\gamma$, where $\delta_0\approx \gamma_0$, satisfy similar correlations 
\begin{equation}
(1+z)\,E_p\propto [E_{iso}]^{1/2},
\end{equation}
while far off-axis ones ($\theta^2\! \gg \! 1/\gamma^2$ and consequently
$\delta_0<<\gamma_0$) have a much lower $E_{iso}$, and satisfy 
\begin{equation}
(1+z)\,E_p\propto [E_{iso}]^{1/3}.
\end{equation}
These  $[E_p,E_{iso}]$ correlations   
that were predicted by the CB model [40]  and later discovered 
empirically [41] for ordinary LGRBs are compared in Figures 3,4
to the observational data on GRBs with known redshift. 
As demonstrated in these figures, 
the CB model correlations predicted for LGRBs,
are well satisfied by both ordinary LGRBs (Eq.(4)) and 
low  luminosity (LL)  LGRBs (Eq.(5)).
The $[E_p,E_{iso}]$ correlation predicted by the CB 
model for LL SGRBs, is also presented in Figure 5. 
As shown in Figure 5, Eqs.(4) (5) seem to be 
satisfied by the observational data on SHBs as well.
\begin{figure}[]
\centering
\epsfig{file=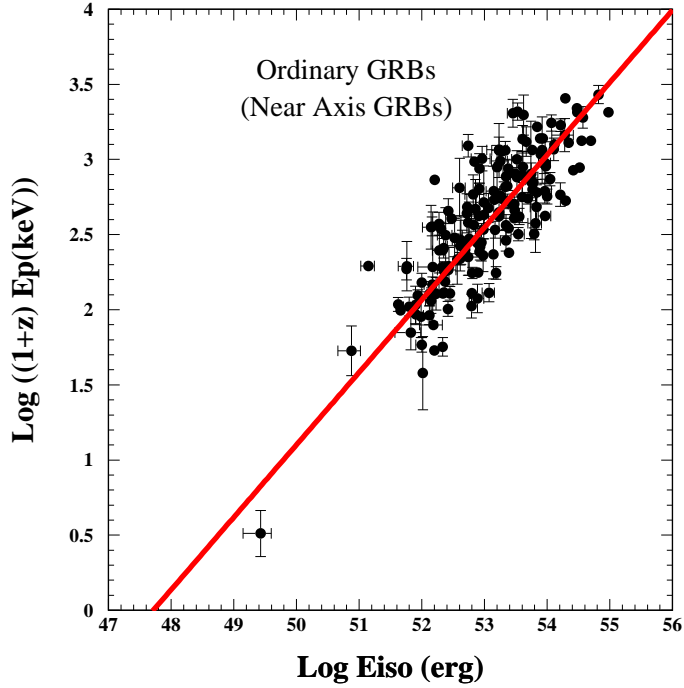,width=10.cm,height=10.cm}
\caption{The $[E_p,E_{iso}]$ correlation in 
ordinary LGRBs viewed near axis. The line is 
the the best fit correlation which is very consistent with the CB model 
predicted correlation given by Eq.(4).}
\end{figure}
\begin{figure}[]
\centering
\epsfig{file=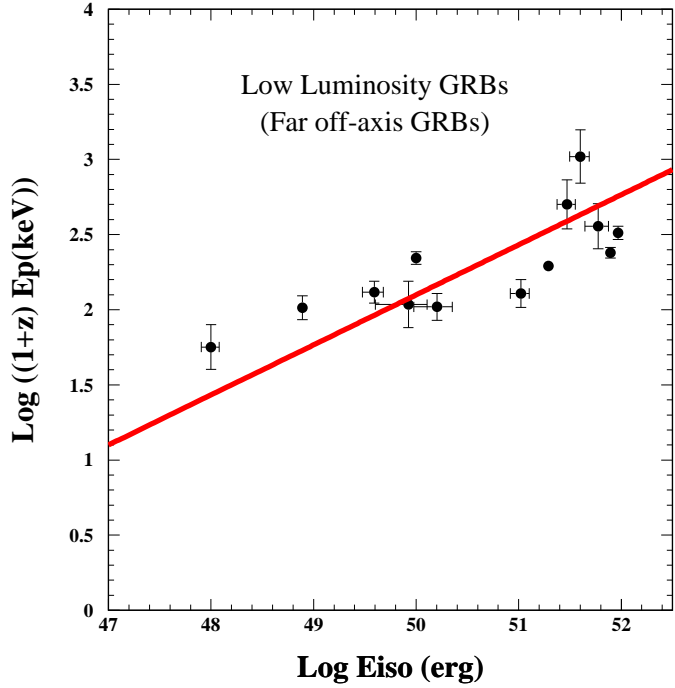,width=10.cm,height=10.cm}
\caption{The $[E_p,E_{iso}]$ correlation in low luminosity 
(far off-axis) LGRBs. The line is the CB model predicted correlation 
as given by Eq.(5).}
\end{figure}
\begin{figure}[]
\centering
\epsfig{file=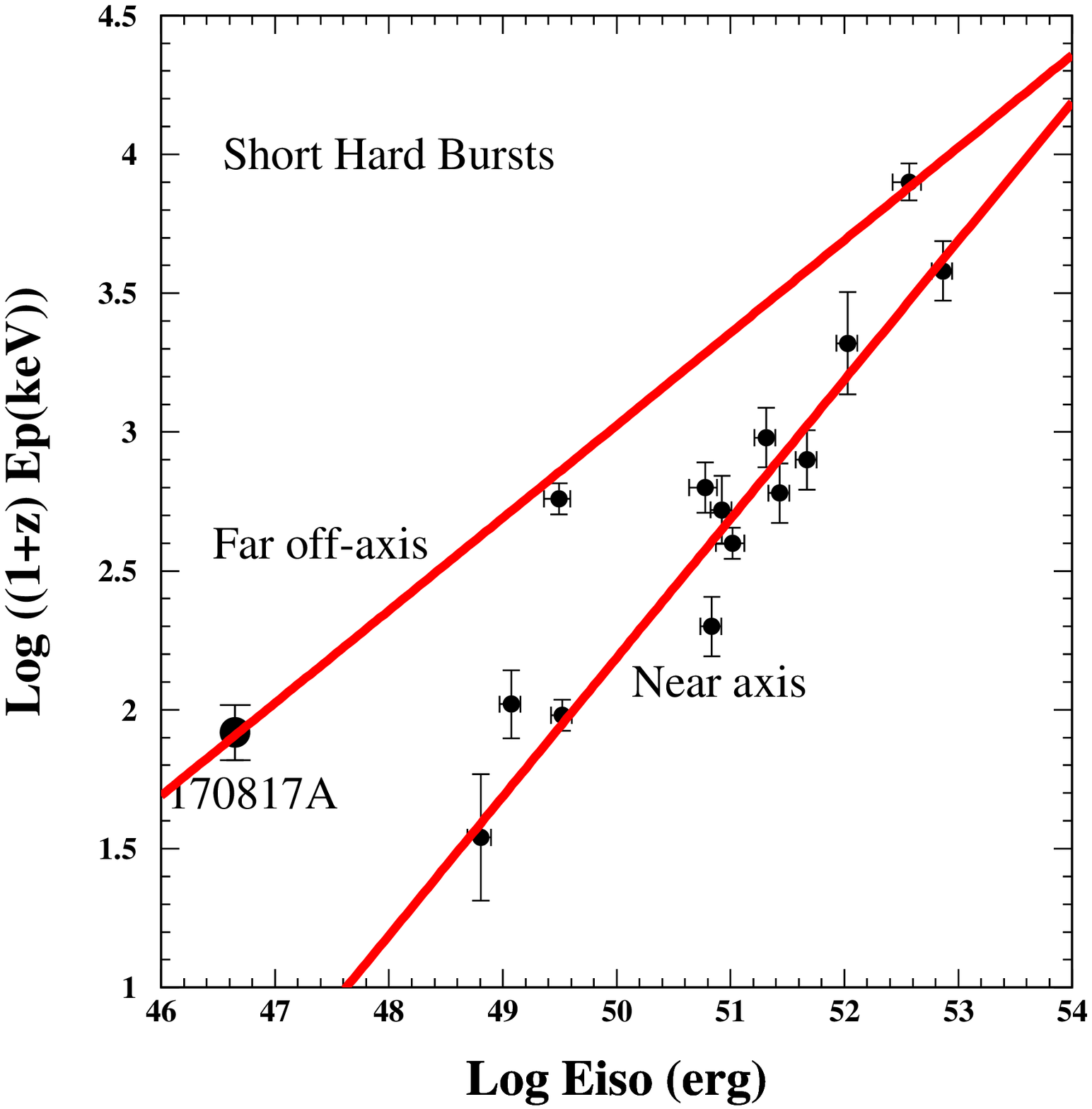,width=10.cm,height=10.cm}
\caption{The $[E_p,E_{iso}]$ correlations in SHBs.
The lines are the CB model predicted correlations as 
given by Eqs (4) and (5).}
\end{figure}

\noindent
{\bf The  FB models,} have not provided,  so far, a plausible 
derivation of the above well established correlations.

\noindent
{\bf Test 3: Pulse Shape}\\
GRBs seem to consist of individual short pulses with roughly a 
fast rise and an exponential decay (FRED) pulse shape [1].
Although the number of such pulses, their 
time sequence, relative intensities and durations that vary 
drastically between bursts, cannot be predicted by the current GRB 
models, the typical FRED shape of individual pulses can
be predicted.

\noindent
{\bf In the CB model}, the pulse-shape produced by ICS of glory 
photons with an exponentially cut off power law (CPL) spectrum, 
$dn_g/d\epsilon\!\propto\! \epsilon^{-\alpha}\,exp(-\epsilon/\epsilon_p)$ 
at redshift $z$, by a CB is given approximately [42] by 
\begin{equation}
E{d^2N_\gamma\over dE\,dt}\!\propto\!{t^2\over(t^2\!+\!\Delta^2)^2}\,
 E^{1-\alpha}\,exp(-E/E_p(t)) 
\label{Eq6} 
\end{equation} 
where $\Delta$ is approximately the peak time of the pulse in the observer 
frame, which occurs when the CB becomes transparent to its internal
radiation, and $E_p\!\approx\! E_p(t\!=\!\Delta)$.
In eq.(6), the early temporal rise like $t^2$ is produced by the  increasing 
cross section, $\pi\, R_{CB}^2\!\propto\! t^2$, of the fast expanding CB when it 
is still opaque to radiation. When the CB 
becomes transparent to radiation due to its fast expansion, its 
effective cross section for ICS becomes a constant equal to  
Thomson cross times the number of electrons in the CB.   
That, and the density $n_g$ of the ambient photons, which for a distance  
$r\!=\!\gamma\,\delta\,c\,t/(1\!+\!z)\!>\! R_g$ (the radius of the 
glory) decreases like 
$n_g(r)\!\approx\!n_g(0)\,(R_g/r)^2\!\propto\! t^{-2}$, produce the temporal 
decline like $t^{-2}$. If  CBs are  launched along the axis of a glory  
of a torus-like pulsar wind nebula, or of an accretion disk with a 
radius $R_g$, then glory photons at a distance $r$  from 
the center  intercept intercept the CB 
at a  lab angle $\theta_{int}$, which satisfies 
$\cos\theta_{int} \!=\!-\! r/\sqrt{r^2\!+\! R_g^2}$. 
It yields a $t$-dependent peak energy,
$E_p(t)\!=\!E_p(0)(1\!-\!t/\sqrt{t^2\!+\!\tau^2})$
with  $\tau\!=\!R\, (1+z)/\gamma\,\delta\,c$, 
and $E_p\!\approx\! E_p(t\!\approx \!\Delta)$, where $\Delta$ 
is approximately the peak time of the pulse.

For LGRBs with $\tau\! \gg \!\Delta$, Eq.(6) yields half maximum 
values at $t\!\approx \!0.41\,\Delta$ and $t\!=\!2.41\,\Delta$, which 
yield a full width at half maximum $FWHM\!=2\,\Delta$, a rise time 
from half maximum to peak value $RT\!=\!0.59\,\Delta$ and a decay time 
from peak count to half peak, $DT\!=\!1.41\,\Delta$. Consequently 
$RT/DT\!\approx\!0.42$ and $RT\approx 0.30\,FWHM$.

The predicted pulse shape as given by Eq.(6) is demonstrated 
in Figure 6 for the single-pulse of GRB930612, which was
measured with BATSE  aboard CGRO. 
\begin{figure}[]
\centering
\epsfig{file=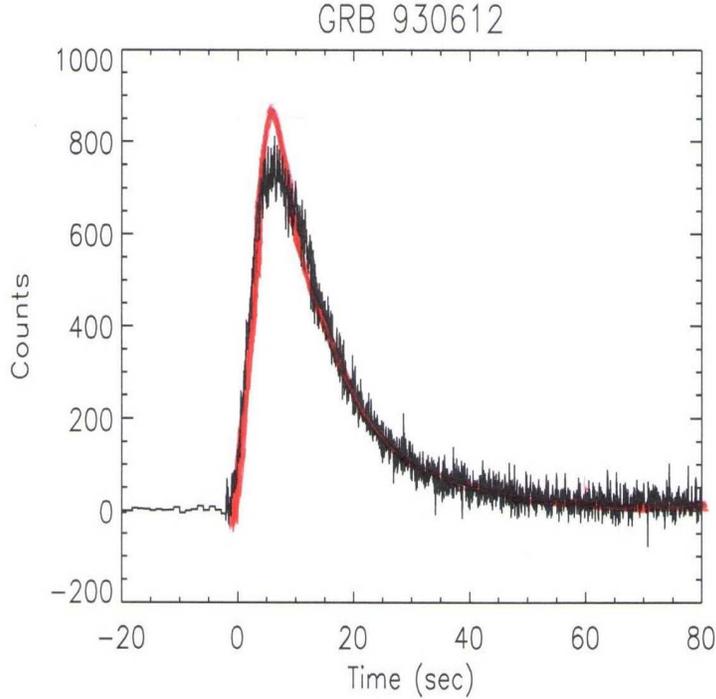,width=10.cm,height=10.cm}
\caption{The pulse shape of GRB930612 measured with BATSE (trigger 2387) aboard 
CGRO, and the shape given by Eq.(6) for the best fit parameters 
$\Delta\!=\!6.2$ s and $\tau\!=\!76.3$ s.}
\end{figure}
In most LGRBs $\tau\!\gg\!\Delta$.
Consequently, the CB model yields for LGRBs 
a pulse asymmetry ratio $RT/DT\!\approx\!0.42$, and $RT/FWHM\!\approx\!0.29$. 
Moreover, these two ratios change very little as long as $\tau\!\gg\!\Delta$. 
Even in the very rare cases where  $\tau/\Delta\!\approx\!1$,
$RT/DT\!\approx\!0.57$ and  $RT/FWHM\!\approx\! 0.36$. 
In Figures 7,8, the CB model predicted ratios  $RT/DT$ and $RT/FWHM$
for $\Delta\!<\!\tau\!<\!\infty$  are compared   
to their best fit values in 77 resolved pulses of
BATSE/CGRO LGRBs reported by Kocevski et al. [43]. 
As shown in Figures 7,8, their best fit values 
lie well within the narrow area between the predicted  CB model boundaries,
and  their mean values $RT/DT\!=\!0.47\pm 0.08$ and $RT/FWHMT\!=\!0.31$ 
reported in [43] are very close to the CB model expected values 
$RT/DT\!=\!0.44$ and $RT/FWHM\!=\!0.31$ for $\tau\!=\! 10\,\Delta$.
\begin{figure}[]
\centering
\epsfig{file=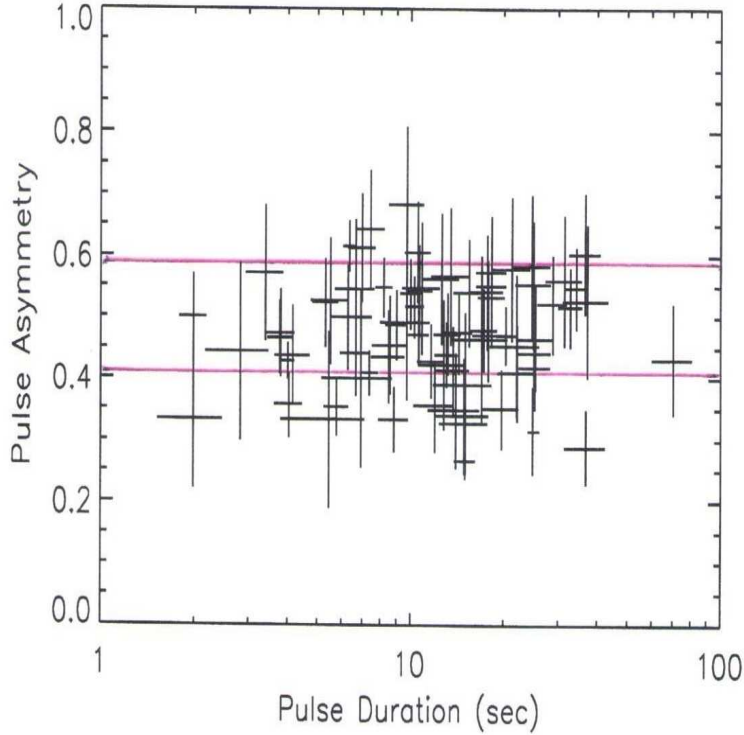,width=10.cm,height=10.cm}
\caption{Comparison between the observed asymmetry ratio $RT/DT$ 
as function of pulse duration reported in [43]
for a sample of 77 resolved LGRB pulses measured with BATSE 
aboard CGRO (with a  mean value $RT/DT\!=\!0.47\pm 0.08$), 
and the CB model predicted asymmetry $0.41\!<\!RT/DT\!<\!0.58$
for $\Delta\!<\!\tau\!<\!\infty$ (solid lines).}
\end{figure} 
\begin{figure}[]
\centering
\epsfig{file=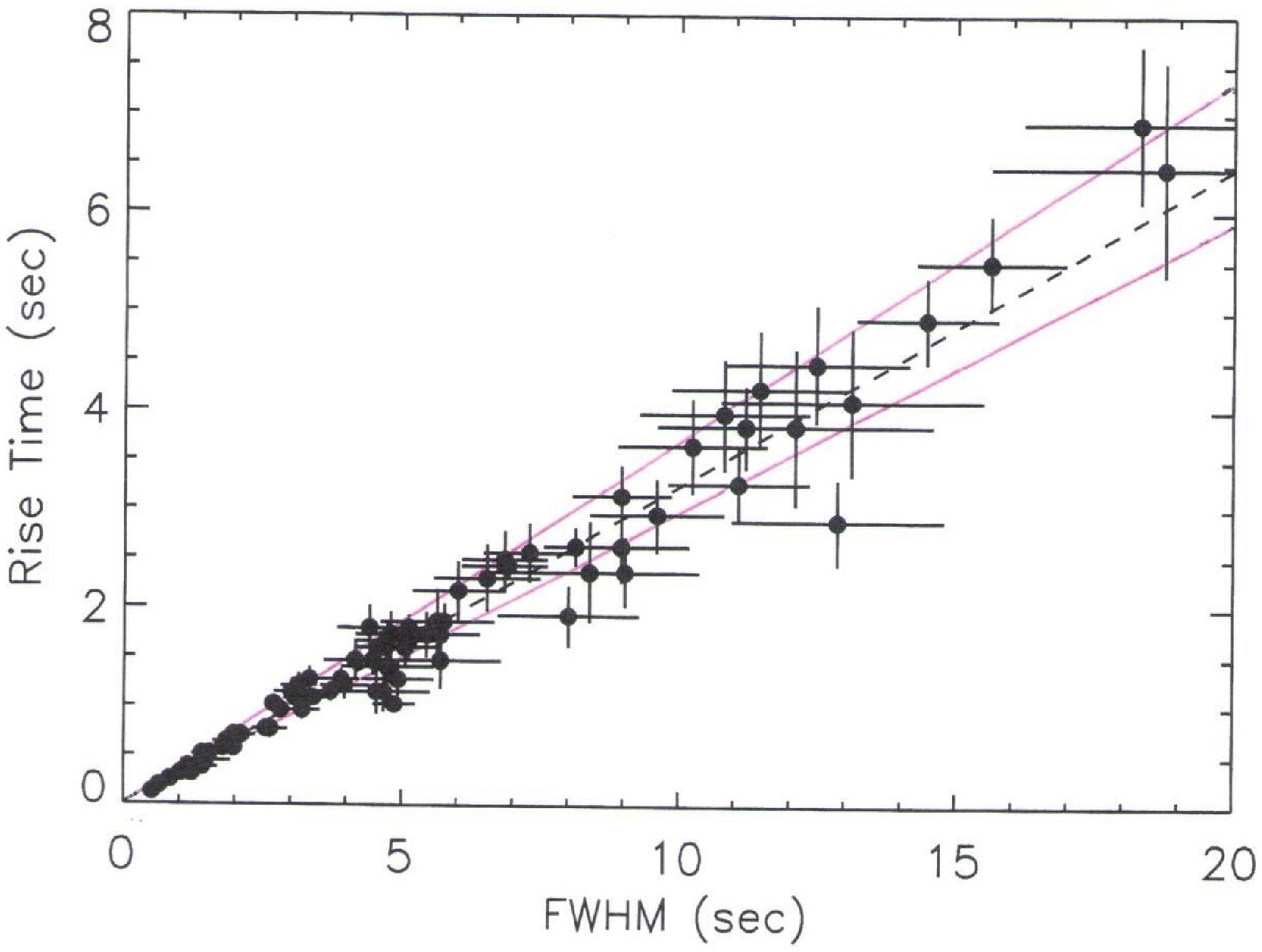,width=10.cm,height=10.cm}
\caption{Comparison between the rise time $RT$ versus the $FWHM$ reported  
in [43] for a sample of 77 resolved pulses measured with BATSE aboard
CGRO. The dotted line is best fit ratio $RT/FWHM\!=\!0.32$ and the solid lines
are CB model expected boundaries  $0.29\!<\! RT/FWHM \!<\!0.36$ for LGRBs.}
\end{figure}
In Figure 9 we compare the measured pulse shape of SHB170817A 
and the CB model pulse shape as given in [42] with  the best 
fit parameters $\Delta\!=\!0.62$ s and $\tau\!=\!0.57$ s
($\chi^2/dof\!=\!0.95$). The best fit light curve has a maximum 
at $t\!=\!0.43$, a  half maximum  at  $t\!=\!0.215$ s
and $t\!=\!0.855$ s, with  an asymmetry $RT/DT\!=0.50$
and $RT/FWHM\!=\!0.34$. 
\begin{figure}[]
\centering
\epsfig{file=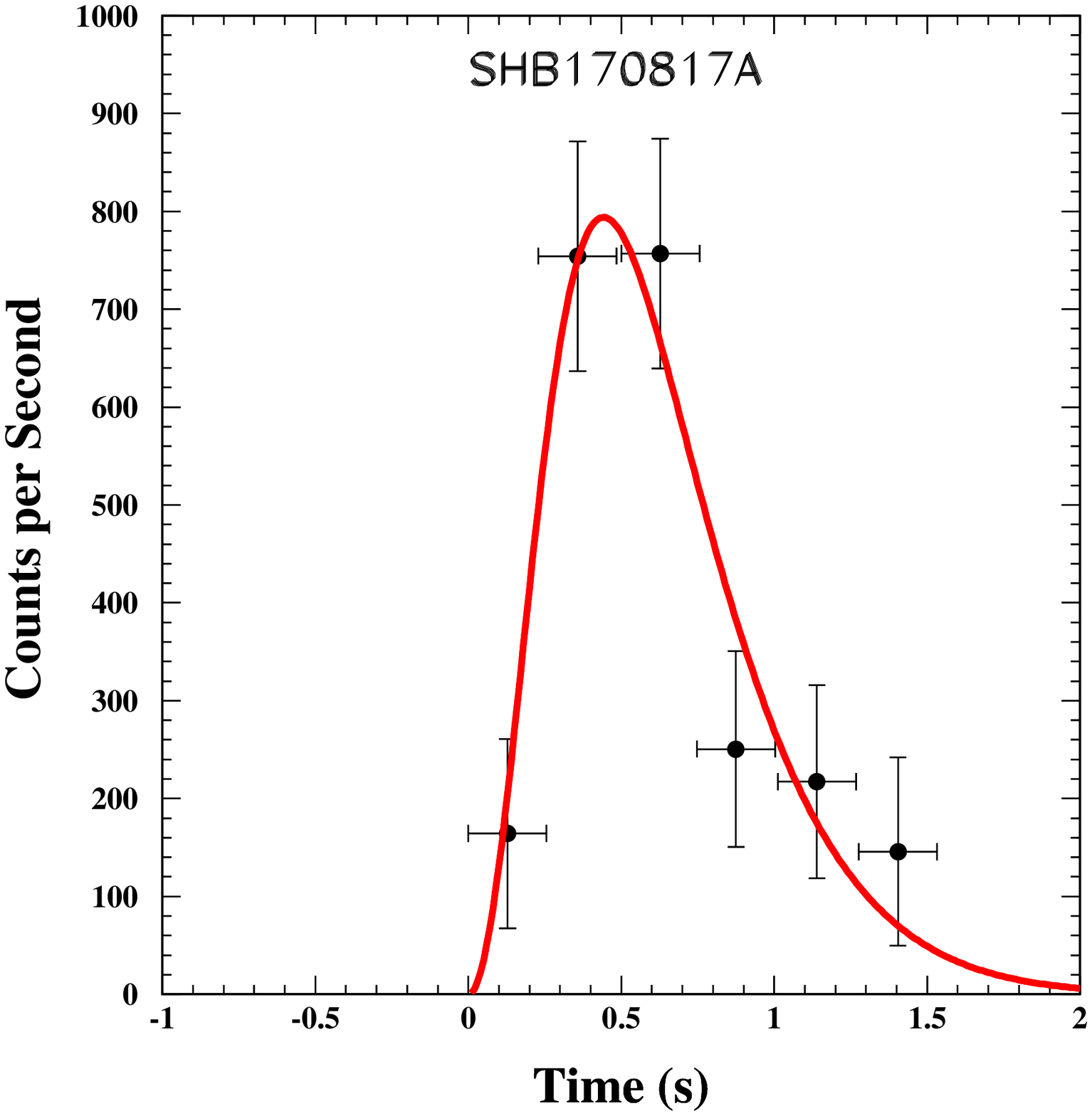,width=10.cm,height=10.cm}
\caption{The pulse shape of SHB170817A measured with the Fermi-GBM
[20] and the best fit pulse shape given by Eq.(6) with 
$\Delta\!=\!0.62$ s and $\tau\!=\!0.57$ s.}
\end{figure}

\noindent
{\bf In the standard fireball models [27]} the GRB prompt emission  pulses are 
produced by synchrotron radiation from shock accelerated electrons in 
collisions between overtaking thin shells ejected by the central engine 
or by internal shocks in the ejected conical jet.
Only for the fast decline phase of the prompt emission, and only in the 
limit of very thin shells and fast cooling, falsifiable predictions were 
derived from the underlying FB model assumptions. 
In this limit the fast 
decline phase of a pulse was derived from the relativistic curvature 
effect [39,44,45]. It yields a power law decay
\begin{equation}
F_\nu(t)\!\propto\!(t\!-\!t_i)^{-(\beta + 2)}\nu^{-\beta}
\end{equation}
where $F_\nu(t)\!=\!E\,dN/dE$,
$t_i$ is the beginning time of the decay phase, 
and $\beta$ is the spectral index of  prompt emission.

The observed exponential decay of the prompt emission accompanied 
by a fast spectral softening before the afterglow took over  
could be  roughly reproduced  by adjusting a beginning time of 
the decay and replacing the constant 
spectral index of the model by the observed time-dependent one [39].
 
\section{Afterglow tests}
{\bf In the CB model}, the afterglow of SN-GRBs 
is mainly synchrotron radiation emitted by the highly relativistic jets 
of CBs launched in core collapse supernova of type Ic (SN-GRBs) in the 
dense interstellar medium (e.g. molecular clouds where most SNeIc of 
short lived massive stars take place).  The afterglow of SN-less GRBs 
and ordinary SHBs seems to be dominated by a PWN emission powered by 
the spin down of a newly born  millisecond pulsar [25,26].

In SN-GRBs, the ionized medium in front of the CBs is 
swept in and generates within them a turbulent magnetic field 
whose energy density is assumed to be in an approximate equipartition 
with that of the swept in particles. The electrons that enter the 
CB with a Lorentz factor $\gamma(t)$ in the CB's rest frame are 
Fermi accelerated there and cool by emission of synchrotron 
radiation (SR), which is isotropic in the CB's rest frame. In the 
observer frame, the emitted photons are beamed into a narrow cone 
of an opening angle $\theta\sim 1/\gamma(t)$ along the CB's 
direction of motion by its highly relativistic motion, their 
arrival times are aberrated, and their energies are boosted by its 
Doppler factor $\delta$ and redshifted by the cosmic expansion 
during their travel time to the observer [28].

The observed spectral energy density (SED) flux of the {\it 
unabsorbed} synchrotron X-rays, $F_\nu(t)=\nu\,dN_\nu/d\nu$, 
has the form (see, e.g., Eqs.~(28)-(30) in [46]),
\begin{equation}
F_{\nu} \propto n^{(\beta_x+1)/2}\,[\gamma(t)]^{3\,\beta_x-1}\,
[\delta(t)]^{\beta_x+3}\, \nu^{-\beta_x}\, ,
\label{Eq8}
\end{equation}
where $n$ is the baryon density of the external medium encountered 
by the CB at a time $t$ and  $\beta_x$ is the spectral index 
of the emitted X-rays, $E\,dn_x/dE\propto E^{-\beta_x}$.

The swept-in ionized material by the CBs decelerates them. 
Energy-momentum conservation for such a plastic collision between 
a CB of a baryon number $N_{_B}$, a radius $R$ and an initial
Lorentz factor $\gamma(0)\gg 1$,
which  propagates in a constant density ISM at a redshift $z$,   
yields the deceleration law in [42]),
\begin{equation}
\gamma(t) = {\gamma_0\over [\sqrt{(1+\theta^2\,\gamma_0^2)^2 +t/t_d}
          - \theta^2\,\gamma_0^2]^{1/2}}\,,
\label{Eq9}
\end{equation}
where $t$ is the time in the observer frame from  
the beginning of the afterglow emission by the CBs,  and 
\begin{equation}
t_d\!=\!{(1\!+\!z)\, N_{_B}/ 8\,c\, n\,\pi\, R^2\,\gamma_0^3} 
\label{Eq10}
\end{equation}
is its  deceleration time-scale. 

In the case of SN-less LGRBs, which  probably are  produced 
by jets ejected in a phase transition of  n*s 
to q*s  in high mass X-ray binaries following mass accretion 
on the n*s [26], the afterglow appears to be dominated by 
radiation emitted by the pulsar's wind nebula,  powered by the rotational energy loss 
of the newly born q*  through magnetic dipole radiation, relativistic 
wind and high energy charged particle emission along open  magnetic lines
[26].     

\noindent
{\bf Test 4: Canonical behavior.}\\ 
{\bf In the CB model,} the prompt $\gamma$-ray emission was 
predicted [46] to end with an exponential 
temporal decay with a fast spectral softening (Eq.6), which is 
taken over by an X-ray afterglow with a shallow decay phase 
("plateau") that breaks smoothly into a power law-decline. This 
"canonical behavior" [47] was predicted by the CB model (see, e.g., 
Figures 26-33 in [46]. Figure 31 is  
shown here as Figure 10) long before the plateau was first observed 
in the X-ray afterglow of GRB050315 [48] and GRB050319 [49], 
with the Swift X-ray Telescope (XRT), as shown in Figure 11. 
\begin{figure}[]
\centering
\epsfig{file=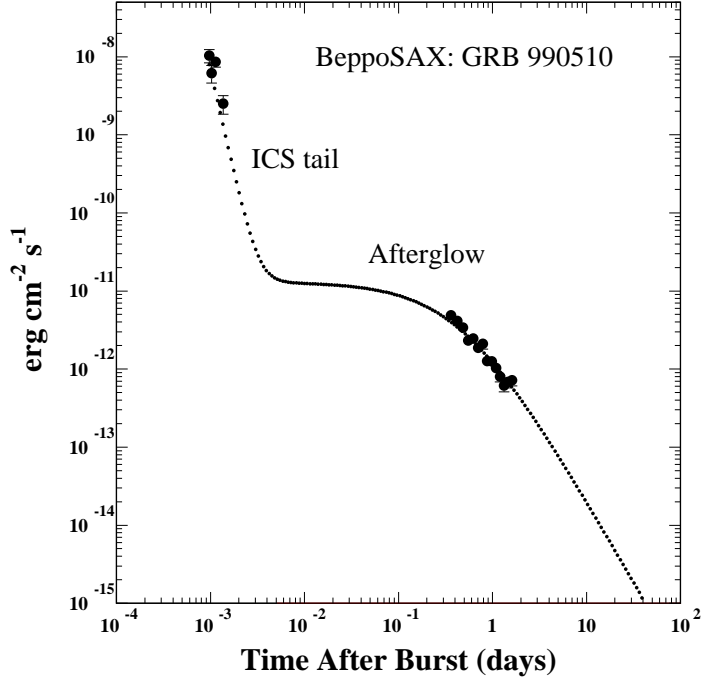,width=10.cm,height=10.cm}
\caption{The X-ray afterglow of GRB 990510 
measured with the telescopes aboard the BeppoSAX satellite
compared to a canonical X-ray afterglow 
predicted by the CB model [46] for SN-GRBs.}
\end{figure}
\begin{figure}[]
\centering
\epsfig{file=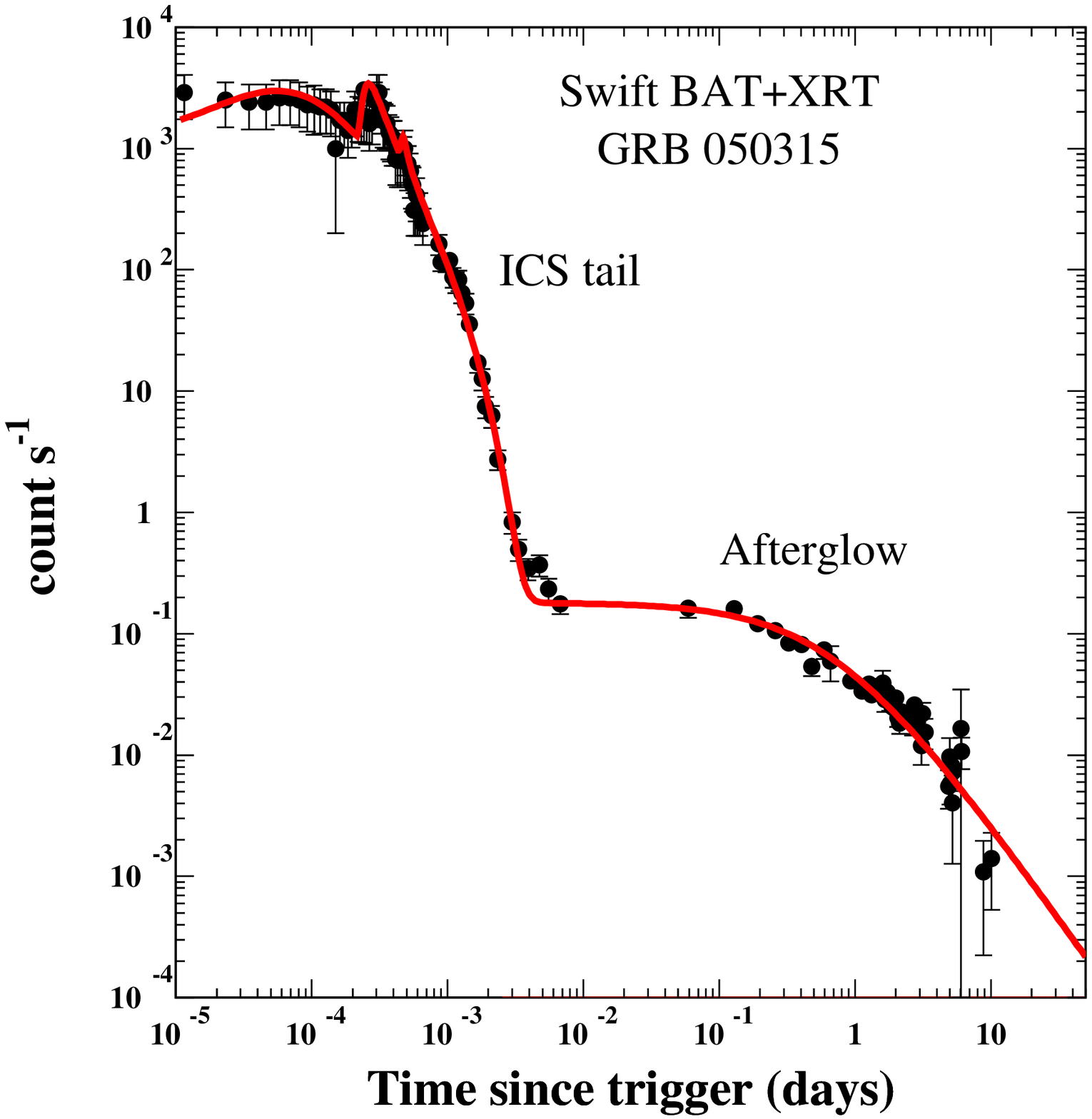,width=10.cm,height=10.cm}
\caption{The X-ray afterglow of GRB050315 
measured with the telescopes aboard  Swift 
compared to its best fit  canonical X-ray afterglow
predicted by the CB model [46] for SN-GRBs.}
\end{figure}

\noindent
{\bf In the FB model,} no canonical behavior is expected.
 
\noindent
{\bf Jet Break Tests}.\\ 
{\bf In the CB model}, the Lorentz factor $\gamma(t)$ of a CB, which decelerates in a 
constant density ISM as given by Eq.(9), change rather slowly as long as 
$t<t_b$, where 
\begin{equation}
t_b\!\approx\! (1\!+\!\gamma_0^2\theta^2)^2 t_d\,. 
\end{equation} 
This slow change produces the plateau phase in SN-GRBs. The explicit 
dependence of $E_p$ and $E_{iso}$ on $\gamma_0$ and $\delta_0$ can be used 
to obtain from Eq.(11) the correlation [50], 
\begin{equation} 
t_b/(1+z)\!\propto\![(1\!+\!z)\,Ep\,Eiso]^{-1/2}, 
\end{equation} 

\noindent
{\bf Test 5: Break time correlations.}\\
The observed break time of the X-ray afterglow of SN-LGRBs
measured with the Swift XRT [51], for SN-LGRBs 
with known redshift, $E_p$ and $E_{iso}$,
is compared in Figure 12 to that predicted by Eq.(12). 
As shown in Figure 12, it is well satisfied by such SN-GRBs.
\begin{figure}[] 
\centering 
\epsfig{file=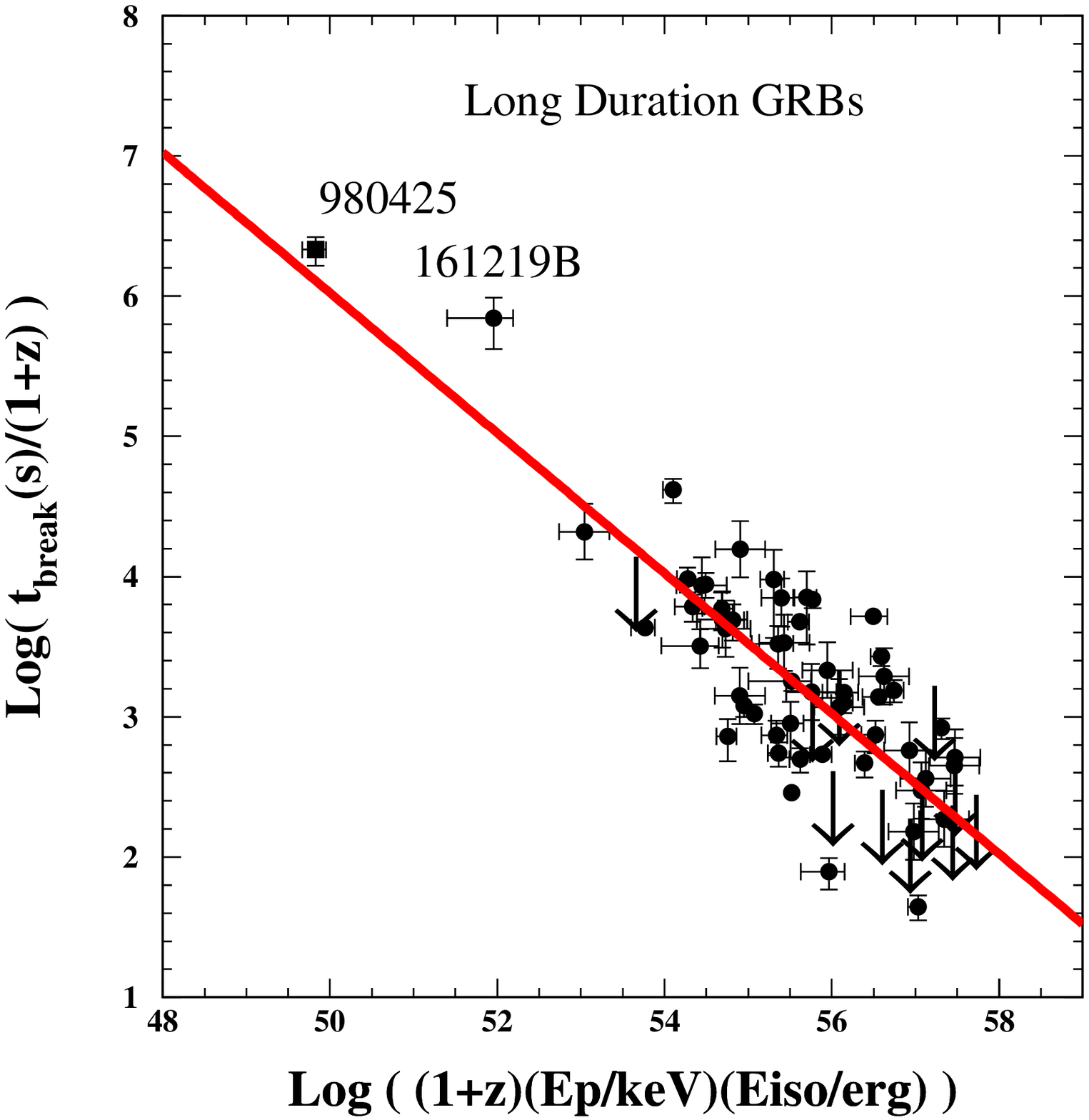,width=10.cm,height=10.cm} 
\caption{The break time $t_b/(1+z)$ of the X-ray afterglow of LGRBs 
measured with the Swift XRT [51], as a function of $[(1\!+\!z)\,Ep\,Eiso]$. 
The line is the CB model correlation given by Eq.(12), which is expected 
in SN-GRBs. SN-Less GRBs that are identified by an afterglow 
with a light curve  $\propto\! 1/(1\!+\!t/t_b)^2$ [26] 
are not included.} 
\end{figure}

\noindent
{\bf Test 6: Post break closure relations.}\\
Far beyond the break, Eq.(9) yields $\delta(t)\approx 2\gamma(t)\propto t^{-1/4}$
[50]. When substituted in Eq.(8), it yields the late-time behavior,
\begin{equation}
F_\nu(t\!>>\!t_b)\!\propto\! t^{-\alpha_\nu}E^{-\beta_\nu},  
\end{equation}
which satisfies  the closure relation
\begin{equation}
\alpha_\nu\!=\!\beta_\nu\!+\!1/2.
\end{equation}
This post break closure relation is well satisfied by the X-ray afterglow 
of SN-GRBs [50] as long as the CB moves within roughly a constant density 
interstellar medium. It is demonstrated in Figure 13 for the X-ray 
afterglow of GRB060729 [51], a canonical afterglow of an SN-GRB. Its long 
followed up and well sampled X-ray afterglow yielded  a best fit 
temporal index $\alpha_x\!=\! 1.46\!\pm\!0.03$, which agrees well with 
$\alpha_x\!=\!\beta_x\!+\!1/2=\!1.49\!\pm\!0.07$ for an observed [51]
$\beta_x\!=\!0.99\!\pm\!0.07$ .
\begin{figure}[]
\centering
\epsfig{file=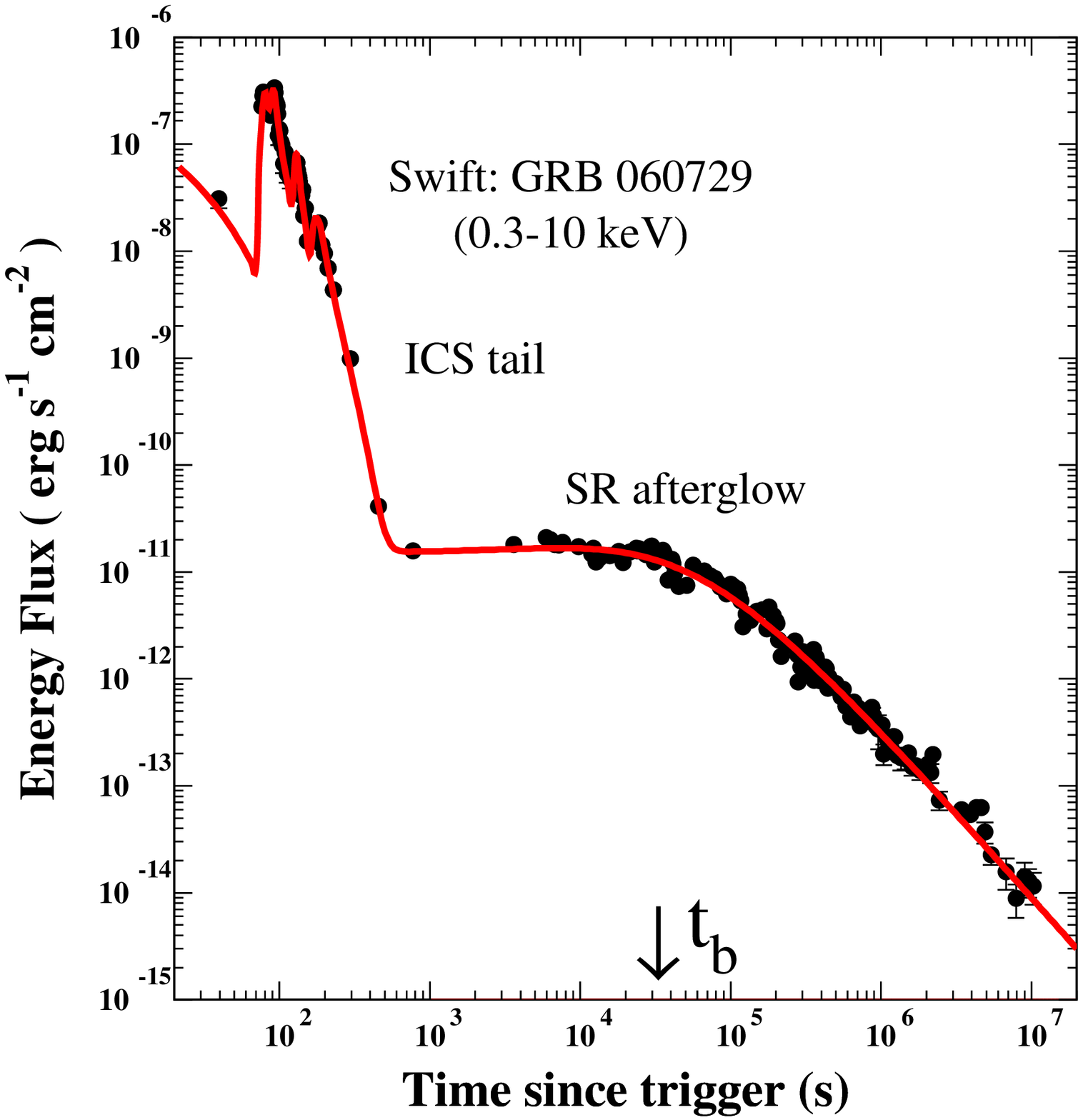,width=10.cm,height=10.cm}
\caption{The canonical light curve of the X-ray afterglow of
the SN-GRB 060729 measured with Swift XRT [51], 
and its best fit CB model afterglow [28] as given by Eq.(8) 
which well satisfies 
the CB model prediction $\alpha_x\!=\!\beta_x\!+\!1/2$.}
\end{figure}  
The most accurate test, however, of the CB model relation  
$\alpha_x\!=\!\beta_x\!+\!1/2$ 
for a single SN-GRB was provided by the follow-up measurements 
of the X-ray afterglow of GRB 130427A, the most intense GRB ever 
detected by Swift and followed with the Swift XRT 
and other sensitive X-ray telescopes aboard  XMM Newton and CXO up to a 
record time of 83 Ms after burst [52]. The measured light-curve  
has a single power-law decline with $\alpha_x\! =\! 1.309\!\pm\! 0.007$ in the 
time interval 47 ks - 83 Ms. 
The best single power-law fit to the combined measurements of the X-ray 
light-curve of GRB 130427A  with the Swift-XRT [17], 
XMM Newton, CXO [52], and MAXI [53] that is 
shown in Figure 14 yields $\alpha_x\!=\!1.294\!\pm\!0.03$.
The CB model prediction as given by Eq.(14) 
with  the measured spectral index $\beta_x\!=\!0.79\!\pm\! 0.03$ [52],
is $\alpha_x\!=\!1.29\!\pm\!0.03$, in remarkable agreement with its best fit 
value.
\begin{figure}[]
\centering
\epsfig{file=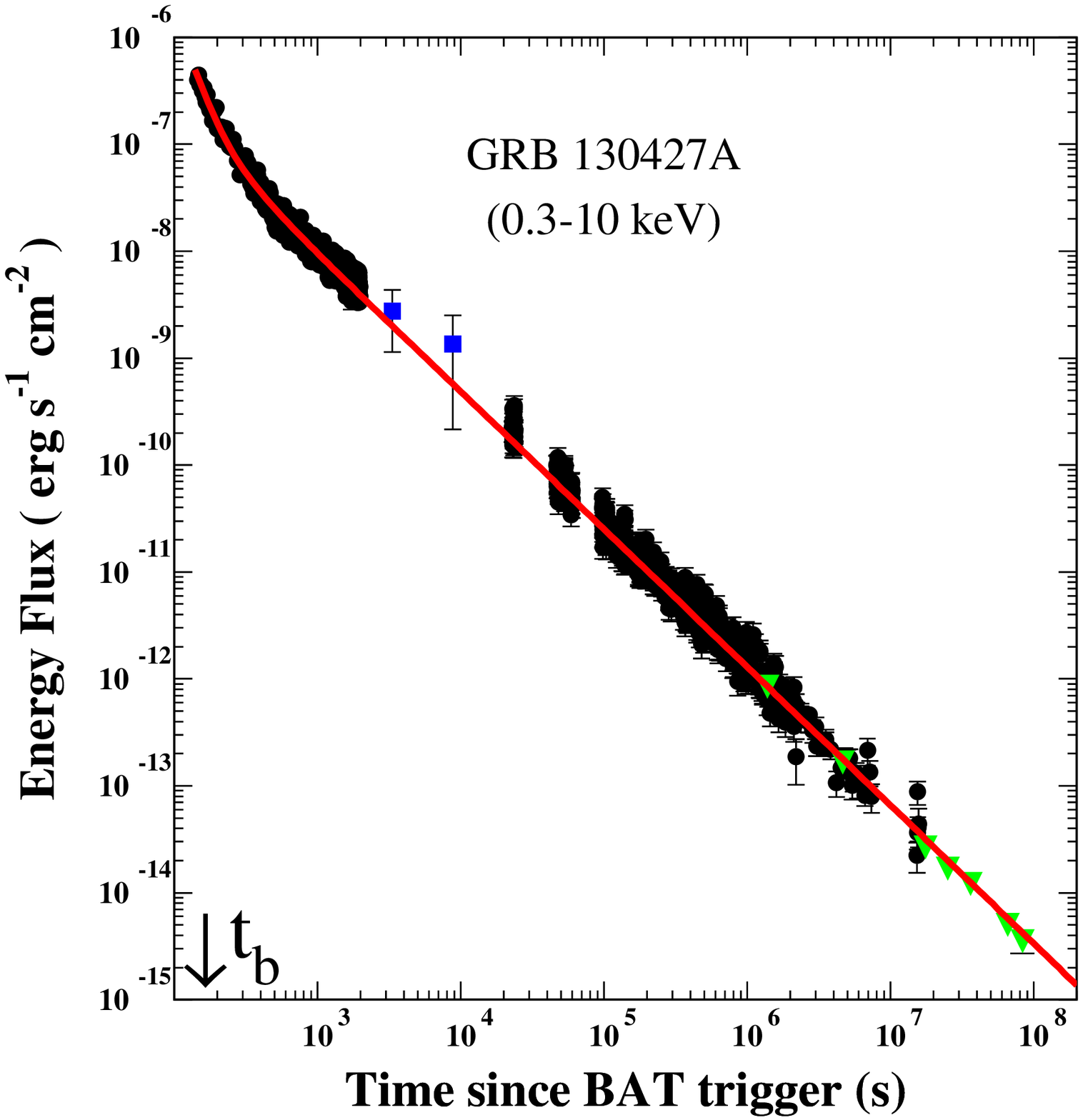,width=10.cm,height=10.cm}
\caption{The X-ray light-curve of the intense GRB 130427A that was
measured with Swift XRT [51] (circles)
and with  XMM Newton  and  Chandra [52] (triangles) up
to 83 Ms after burst, and its CB model best-fit with a start time and an
early break hidden under the prompt emission phase.
Also shown are the two MAXI data points [53] (squares)
at t = 3257 s and t = 8821 s. The best-fit 
power-law decline has an index $\alpha_x=1.29$. The temporal decay index 
predicted  by the CB model, Eq.(14), for the measured spectral index [52] 
$\beta_x\!=\!0.79\!\pm\!0.03$  is  $\alpha_x\! =\!1.29\!\pm\! 0.03$.} 
\label{Fig14}
\end{figure}
No doubt, the assumptions of a constant density circumburst medium 
is an over simplification: SN-LGRBs that are produced 
by supernova explosions of type Ic of short-lived massive stars, 
take place mostly in superbubbles formed by star formation.
Such superbubble environments may have a bumpy density, 
which deviates significantly from the assumed  constant-density ISM.
It may be  responsible for the observed deviations from the predicted 
smooth light-curves and for $\chi^2/dof$ values slightly larger than 1.  
Moreover, in a constant-density ISM, 
the late-time distance of a CB from its launch site 
is given roughly by,  
\begin{equation}
x\!=\!{2 c \int^t \gamma\delta dt\over 1+z}\!\approx\! 
{4\,c\,\gamma_0^2\sqrt{t_b\,t_d}\over {1+z}}\,.    
\end{equation}
This distance can exceed the size of the superbubble and even the scale-height of 
the disk of the GRB host galaxy. In such cases, the transition of a CB from the 
superbubble into the Galactic ISM or into the Galactic halo, in face-on disk 
galaxies, will bend the late-time single power-law decline into a more rapid 
decline, depending on the density profile above the disk. For instance,
when the CB exits the disk into the halo, its Lorentz and Doppler factors 
tend to constant values while its afterglow decays roughly like ((see Eq.(8))
\begin{equation}
F_\nu(t)\!\propto\![n(r)]^{(1\!+\!\beta_\nu)/2}
\end{equation}
where $r\!\propto \!t$. 
Such a behavior may have been observed by the Swift XRT [51] in several 
GRBs, such as 080319B and 110918A, at $t\!>\! 3\times 10^6$ s and in GRB 060729 at 
$t>3\times 10^7$ s by CXO [54]. 

\noindent
{\bf Test 7: Missing breaks.} \\
{\bf In the CB model}, Eq.(8) yields a single power-law for the 
temporal decline of the light curve of the afterglow well beyond the break 
time $t_b$ as long as the CB moves in a constant density interstellar 
medium.
Consequently, in the CB model, very energetic LGRBs, i.e., 
those with a large product $(1\!+\!z)E_pE_{iso}$, may have a break time 
$t_b$ smaller than the time when the afterglow takes over the prompt 
emission, or before the afterglow observations began [55]. In such cases 
the observed afterglow light curve has a single power-law behavior with a 
temporal decay index $\alpha_\nu\!=\!\beta_\nu\!+\!1/2$ and a "missing 
break". This was first observed in GRB 061007 [56], with 
the Swift XRT [51] as demonstrated in Figure 15.
\begin{figure}[]
\centering
\epsfig{file=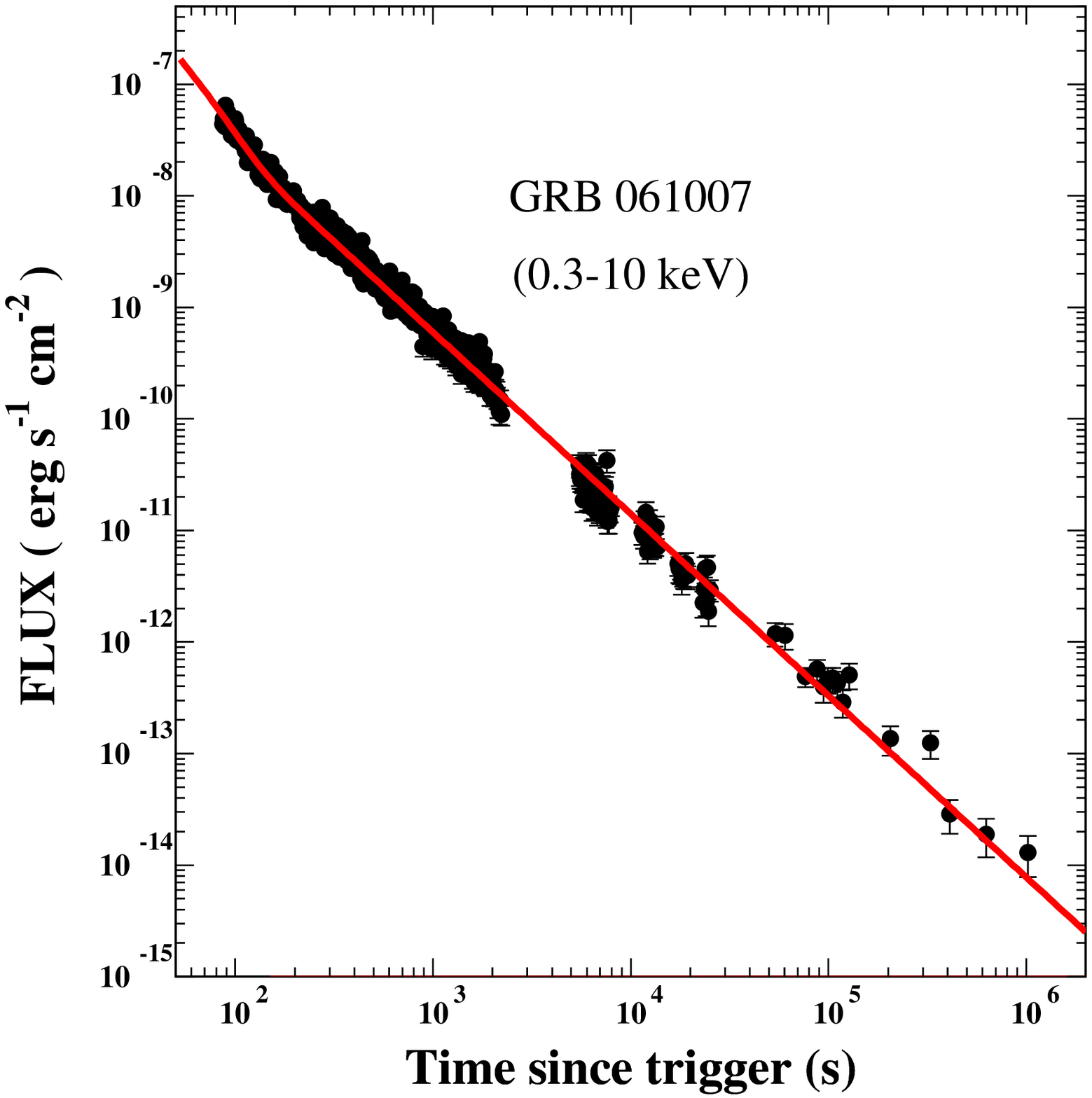,width=10.0cm,height=10.0cm}
\caption{The single power-law best fit to the afterglow of GRB061007 with  
a "missing jet break" measured with Swift XRT [56]. The best fit temporal 
index $\alpha_x\!=\!1.65\!\pm\!0.01$ satisfies the CB model prediction 
$\alpha_x\!=\!\beta_x\!+\!1/2\!=1.60\!\pm\! 0.11$.}
\label{Fig15}
\end{figure}
The $\alpha_x$ values of the most energetic LGRBs with known redshift that 
were obtained from the Swift XRT measurements are plotted in Figure 16 as 
function of their measured $\beta_x+1/2$ values. Also plotted is the best 
fit linear relation $\alpha_x=a\,(\beta_x+1/2)$, which yields a=1.007, in 
good agreement with a=1 predicted by the CB model.
\begin{figure}[]
\centering
\epsfig{file=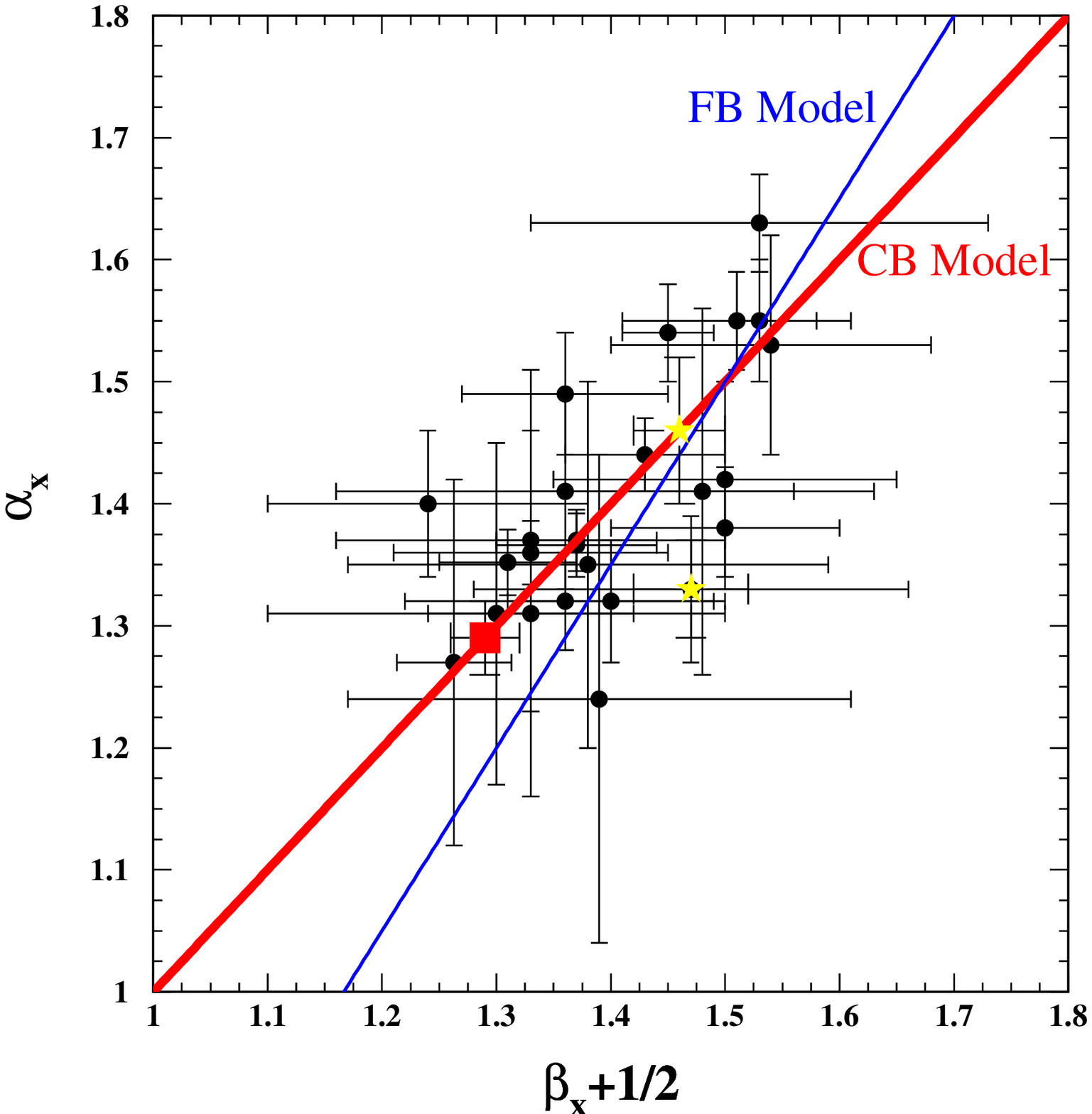,width=10.0cm,height=10.0cm} 
\caption{The values of the post break temporal index $\alpha_x$ as 
function of the spectral index $\beta_x+1/2$ for the 28 most intense GRBs 
with known redshift [57]  that were obtained from 
the follow-up measurements of their 0.3-10 keV X-ray afterglow with
the Swift XRT [51]. The square indicates the value obtained for GRB 130427A.
The thick line is the CB model prediction, Eq.(14).}
\label{Fig16}
\end{figure}

\noindent 
{\bf In the FB models}, the existence of a GRB afterglow at  
longer wave lengths was predicted [12] long before 
it  was discovered in the X-ray band by the telescopes aboard the 
Beppo-SAX satellite [13] and then  by ground based optical 
and radio telescopes [14]. In the first two years 
after their discovery, the observed light curves of these afterglows
were claimed to be well fitted by a single power-law [58],
predicted for  spherical fireballs [59],[60].  Later  when the observations  
clearly indicated a smoothly broken power-law behavior  rather than  
a single power-law, the spherical $e^+e^-\gamma$ fireball have been
replaced without much ceremony, first with conical flows 
or a succession of thin conical shells of $e^+e^-\gamma$ plasma,
and later by plasma of ordinary matter. These flows retained  the  name
 "collimated fireballs"  and the revised models retained the 
"fireball model" name. However, these collimated fireballs could neither 
explain, nor reproduce correctly the observed behavior of the afterglow of 
SN-GRBs and failed tests 4-7:

{\bf Test 4 (Canonical behavior)}: The afterglows were predicted to 
decay like a broken power-law [61],
but could not  explain/reproduce the  'plateau phase' of the afterglow  
observed in many GRBs without postulating a continued energization [62] 
of the jet by the central engine. 

{\bf Test 5 (Break time correlation)}:
In the standard fireball models, the opening angle of the conical jet satisfies 
$\theta_j\!\gg \! 1/\gamma_0$.  Because of relativistic beaming, initially only 
a fraction $\sim\!1/\gamma^2\, \theta_j^2 $ of the front surface of the jet is 
visible to a near axis, distant observer. This fraction, increases with time 
like $[\gamma(t)]^{-2}$, due to the deceleration of the jet in the interstellar 
medium (ISM), until the entire front surface of the jet becomes visible, i.e., 
until $t\!\approx \! t_b$ where $\gamma(t_b\!)=\!=\!1/ \theta_j$.
If the total  $\gamma$-ray energy $E_{\gamma}$  is assumed to be a constant 
fraction $\eta$ of the initial  kinetic energy $E_k$ of the jet, which decelerates 
in an ISM  of a constant baryon density $n_b$  by sweeping in 
({\it plastic collision}) the interstellar matter on its trajectory, then [63]  
\begin{equation}
t_b/(1\!+\!z)\approx {1\over 16\,c}\,\left[{3\,E_{iso}\over \eta\,\pi\, n_b\, 
m_p\,c^2}\right]^{1/3}\,[\theta_j]^{8/3}.
\end{equation}
Although the standard fireball model assumes that the afterglow is 
synchrotron radiation emitted by the shocked ISM (i.e., 
through  {\it elastic scattering} of the ISM particles in front of the jet, 
and not by  {\it plasic collision}, see Table II), Eq.(17)  has been used widely
in the literature to estimate $\theta_j$  without any justification..

Moreover, if $E_\gamma\!\approx\! \eta\, E_k\!\approx\!E_{iso}\theta_j^2/4$ is a
"standard candle" [64], then  $E_{iso}\,\theta_j^2\!\approx\! const$ and
\begin{equation}
t_b/(1\!+\!z)\!\propto\! [E_{iso}]^{-1}.
\end{equation}
The same $[t_b,E_{iso}]$ correlation is obtained for the deceleration of a conical jet 
in a wind-like circumburst density [65]. However, Eq.(18), the expected $[t_b,E_{iso}]$
correlation in the FB model, was shown [66] to be  inconsistent with the best fit correlation
$t_b/(1\!+\!z)\propto E_{iso}^{-0.69\!\pm\!0.06}$ to the observational data. 
The  observational data, however, is consistent with the correlation 
$t_b/(1\!+\!z)\propto E_{iso}^{-0.75}$ expected in the CB model [66].

\begin{table*}
\caption{The late time t-dependence of the bulk motion Lorentz factor 
of highly relativistic conical jets which decelerate by
collision with the surrounding medium.} 
\centering
\begin{tabular}{c c c c c c c}
\hline \hline
Collision:~~~ & Plastic & Plastic  & & Elastic & Elastic\\
\hline 
Density:~~~~~ & ISM &  Wind  &~~~~~ & ISM & Wind\\
\hline\hline
$\gamma(t)\!\propto$& $t^{-3/8}$ & $t^{-1/4}$&~~~~~ &
 $t^{-3/7}$ & $t^{-1/3}$ \\
\hline 
\end{tabular}
\end{table*}
{\bf Test 6: Closure relations.} In the conical fireball model, the 
increase of the visible area of the  conical jet with time  
like $1/[\gamma(t)]^2$  that  stops at $t_b$ yields an achromatic 
temporal break in the GRB  afterglow. If the afterglow is  parametrized as
$F_\nu(t)\!\propto t^{-\alpha}\nu^{-\beta}$, the predicted change in 
$\alpha$ across the break is achromatic and satisfies
\begin{equation}
\Delta(\alpha)\!=\! \alpha(t\!>\!t_b)\!-\!\alpha(t\!<\!t_b)\!=\! 3/4
\end{equation}
for a  constant ISM density. For a wind like 
density, $\Delta(\alpha)\!=\!1/2$. 
This closure relation for either an ISM or a wind like 
density is not satisfied by most GRB breaks, as can also be seen in  
Figures 10-15. In fact, Liang, et al. [67]  analyzed the afterglow of 
179 GRBs detected by Swift between January 2005 and January 2007 and 
the optical AG of 57 pre-Swift GRBs. They did not find any afterglow 
with a break satisfying tests 5,6.\\

\noindent
{\bf Test 7 (Missing breaks):} The missing break in the X-ray 
afterglow of several GRBs with a long follow up measurements was 
suggested to take place after the observations ended [68]. 
But, Eq.(18) implies  that late-time breaks are present only in GRBs with a small 
$E_{iso}$. This, suggestion is in contradiction with the fact that 
missing breaks in GRBs with well sampled afterglows, which extend to 
late times, are limited to GRBs with very large $E_{iso}$, rather 
than small $E_{iso}$. This is demonstrated in Figures 15,14 by the 
unbroken power-law X-ray afterglows of 
GRBs 061007 and 130427A, where 
$E_{iso}\!=\!1E54$ erg [56] and $E_{iso}\!=\!8.5E53$ 
erg [52], respectively, which satisfy well the 
CB model post break closure relation  given by Eq.(14)\\.

\noindent
{\bf Test 8: Chromatic Jet Breaks}\\
{\bf In the CB model} the jet deceleration break in the afterglow of jetted SN-GRBs is dynamic 
in origin and usually chromatic [28].\\
{\bf In the FB model} the jet breaks in the afterglows of jetted SN-GRBs are basically  
geometrical in origin and therefore are predicted to be mostly achromatic 
[27], in conflict 
with observations.
   
\noindent
{\bf Test 9: Universal afterglow of SN-less GRBs.}\\
Figure 17, adapted from [69]
shows the X-ray light curve of GRB 990510 measured with BeppoSAX,
whose afterglow  could not be fit well by a single power-law   
predicted by spherical fireball models (e.g., [69]).    
It could, however, be fit well by an achromatic "smoothly broken 
power law" parametrizations [69] as shown in Figure 18. That, and the observed 
optical and X-ray afterglows of a couple 
of other GRBs detected by BeppoSAX, which could  be fit  by 
a smoothly broken power-law, led to  the replacement of the spherical 
$e^+e^-\gamma$ fireball  by an  $e^+e^-\gamma$  "conical fireball"
which was later replaced with a conical jet of ordinary matter and  
became the  current standard collimated fireball model of GRBs.
\begin{figure}[]
\centering
\epsfig{file=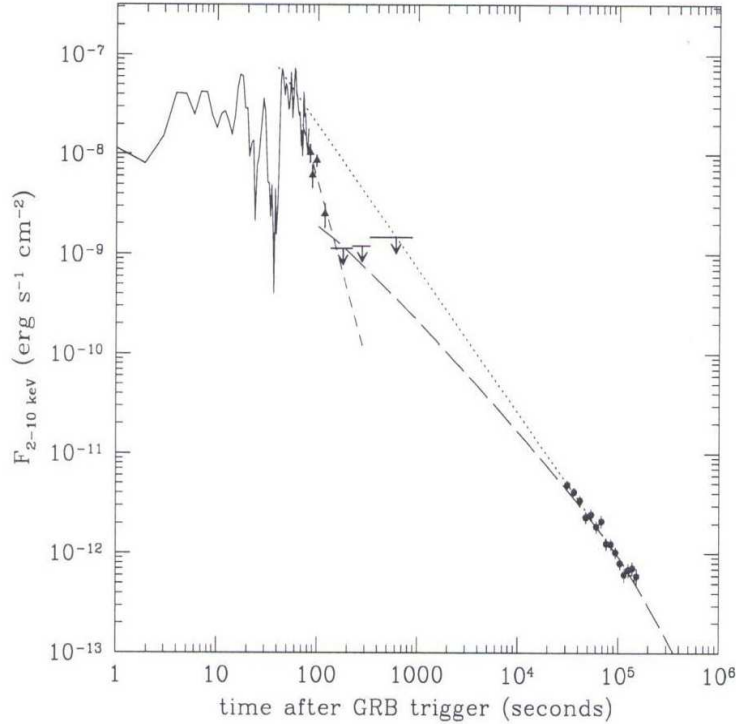 ,width=10.0cm,height=10.0cm}
\caption{The X-ray light curve of GRB 990510 and its  afterglow 
measured with BeppoSAX 
together with a single power-law fit and a smoothly broken single  power-law fit 
to its X-ray afterglow [69].}
\end{figure}
\noindent
But, the afterglow of GRB 990510 and other GRBs which were fit 
by smoothly broken single power laws are not 
conclusive evidence of being produced by a conical jets.  
In fact, an isotropic radiation 
from a pulsar wind nebula powered by a newly born millisecond pulsar 
has an expected luminosity [26] which satisfies 
\begin{equation}
L(t)\!=\! L(0)/(1\!+\!t/t_b)^2
\end{equation}
where $t_b\!=\!P(0)/2\dot{P}(0)$, and  $P(0)$ and $\dot{P}(0)$ 
are, respectively,  the initial period and its time derivative, 
of the newly born pulsar. This is shown in Figure 18 
for GRB 990510. 
\begin{figure}[]
\centering   
\epsfig{file=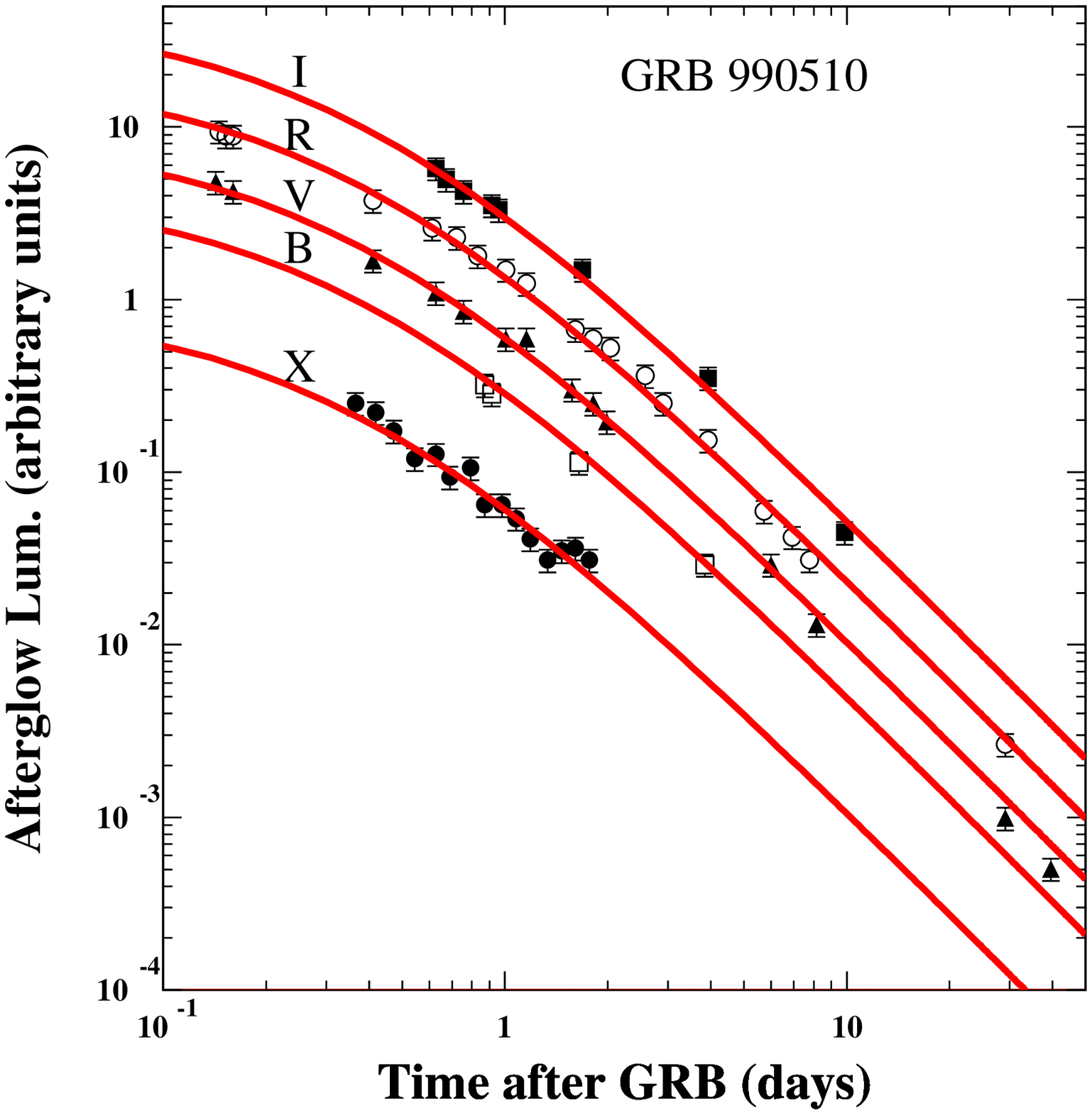,width=10.0cm,height=10.0cm}
\caption{Comparison of Eq.(20), the predicted temporal behavior 
of the light curves of the X-ray and optical afterglows of GRB990510
and their observed light curves. The X-ray data at 5 keV (filled circles) 
is from [69]. The data in the bands 
I (filled squares), R (empty circles), V (filled triangles)
and B (open squares) are that compiled in [69] from [70]. 
The flux normalization is in arbitrary units.}
\end{figure}
In particular, if the afterglows of  SN-less GRBs 
are  produced by PWNs  powered by the newly born MSPs, then $L(t)/L(0)$  
has {\it a universal temporal behavior} [22] as a function of $t/t_b$,  
\begin{equation}
L(t/t_b)/L(0)\!=1/(1\!+\!t/t_b)^2
\end{equation}
This  {\it universal behavior} describes  well the X-ray and optical 
afterglow light curves of GRB 990510, as shown in Figures 18,19.
It also describes well the afterglow of all  SN-less GRBs and SHBs with a well 
sampled afterglow within the first couple of days after burst. 
This is demonstrated  in Figure 20 for the X-ray afterglow of 12  
SN-less GRBs, and in Figure 21 for all the 12 SHBs [22]
from the Swift XRT light curve repository [51] 
with a well sampled afterglow during the first few days 
after burst. 
 \begin{figure}[]
\centering
\epsfig{file=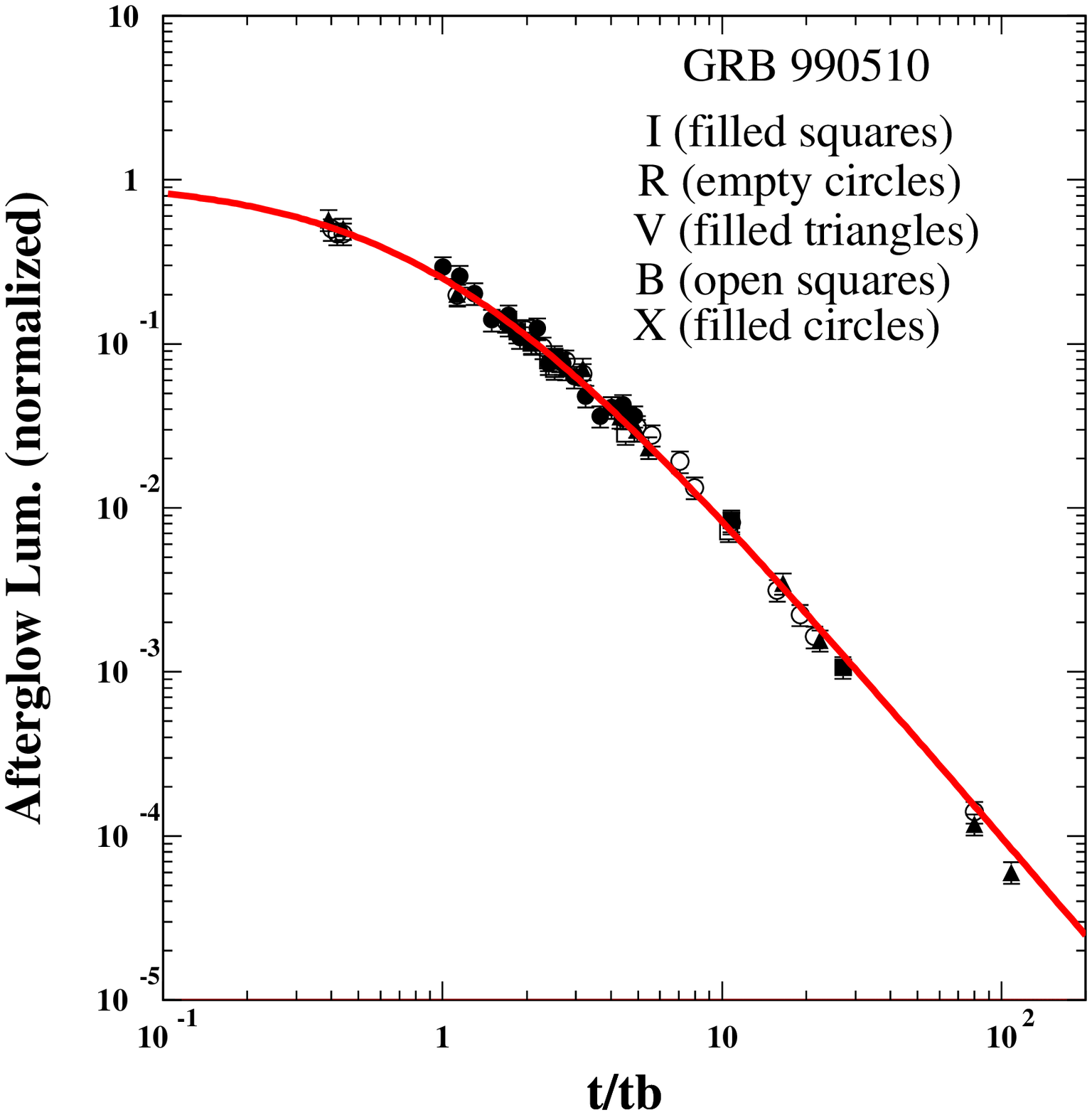 ,width=10.0cm,height=10.0cm}
\caption{Comparison between the normalized  light curves of the X-ray 
and optical afterglows of GRB990510
and their predicted universal shape as given by Eq.(21). 
Data is the same as in Figure 18.}
\end{figure}
\begin{figure}[]
\centering
\epsfig{file=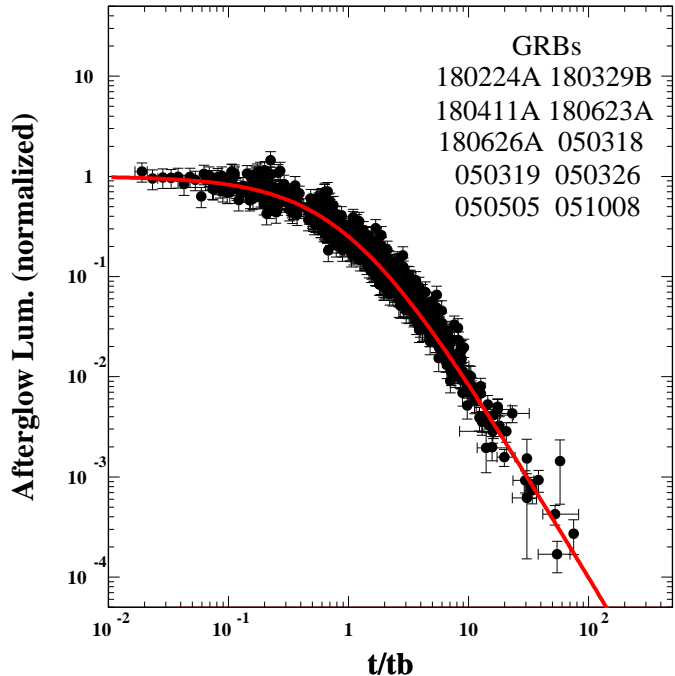,width=10.0cm,height=10.0cm}
\caption{Comparison between the normalized light curve
of the X-ray afterglow measured with Swift XRT [51] 
of 12 SN-less GRBs with a well sampled afterglow
in the first couple of  days after burst
and their predicted universal behavior as given by Eq.(21).}
\end{figure}
\begin{figure}[]
\centering
\epsfig{file=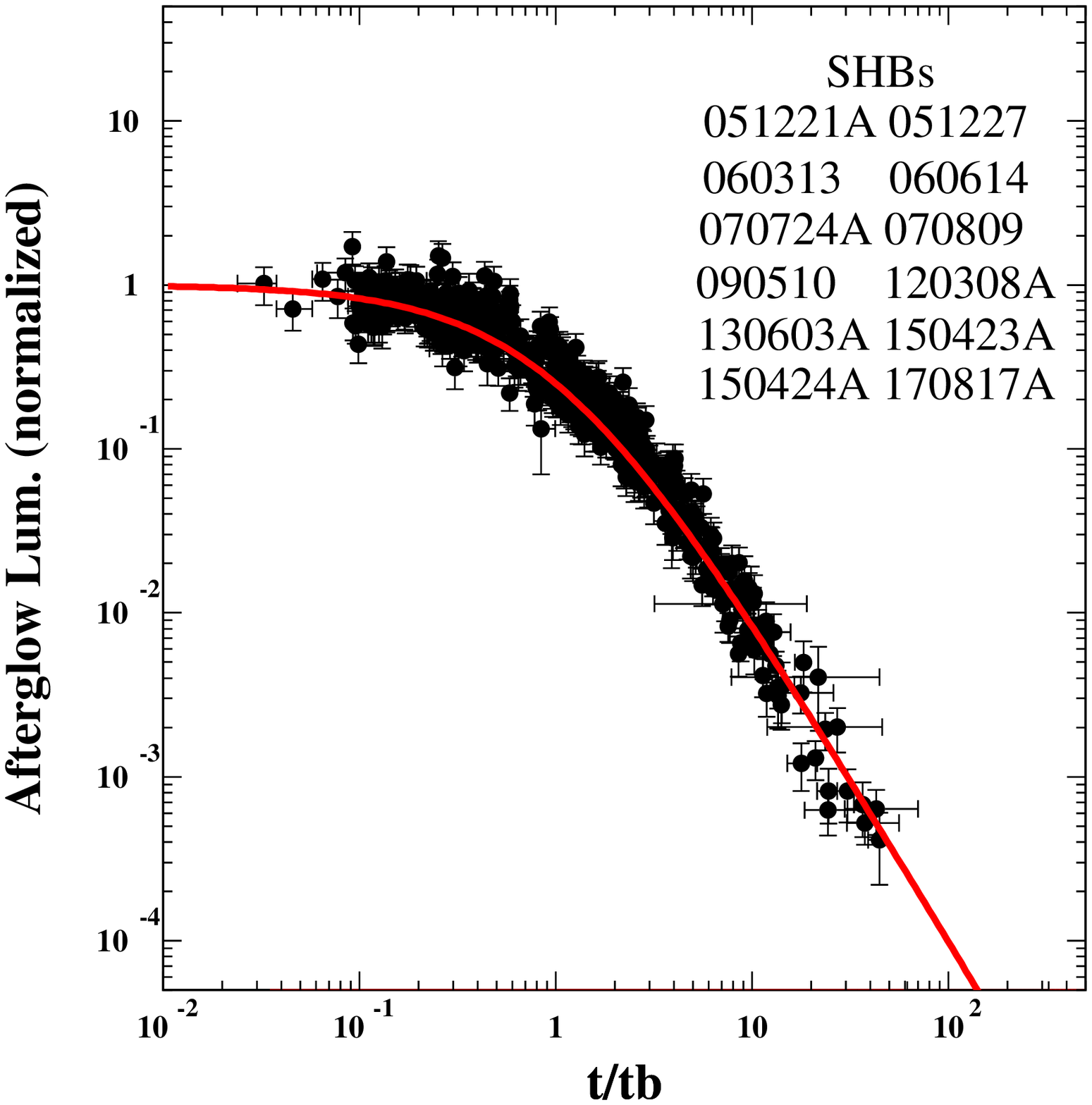,width=10.0cm,height=10.0cm}
\caption{Comparison between the normalized light curve
of the X-ray afterglow of 11 SHBs 
with a well sampled afterglow measured with the Swift XRT [51]
during the first couple of days after burst 
and the predicted universal behavior given by Eq.(21).
The bolometric light curve of SHB170817A reported in [71],
is also included.} 
\end{figure}

\section{Progenitors of long GRBs}
\noindent 
{\bf Test 10: Redshift Distribution of long GRBs}.\\
{\bf In the CB model} long duration GRBs belong to two classes, 
SN-GRBs that are produced in stripped envelope supernovae of type Ic 
(SNeIc) and SN-less GRBs that presumably are produced in a phase 
transition of neutron stars to quark stars following mass accretion in 
high mass X-ray binaries [11,22].
In both cases, the progenitors involve a short lived
high mass star. Hence, in the CB model, the observed production rate of 
long GRBs is proportional to the star formation rate [72] modified by 
beaming [73].\\

Figure 22 compares the observed distribution of long GRBs 
as a function of redshift and their expected distribution in the CB model 
assuming their production rate is proportional to the SFR  modified by 
beaming [73]).
\begin{figure}[]
\centering
\epsfig{file=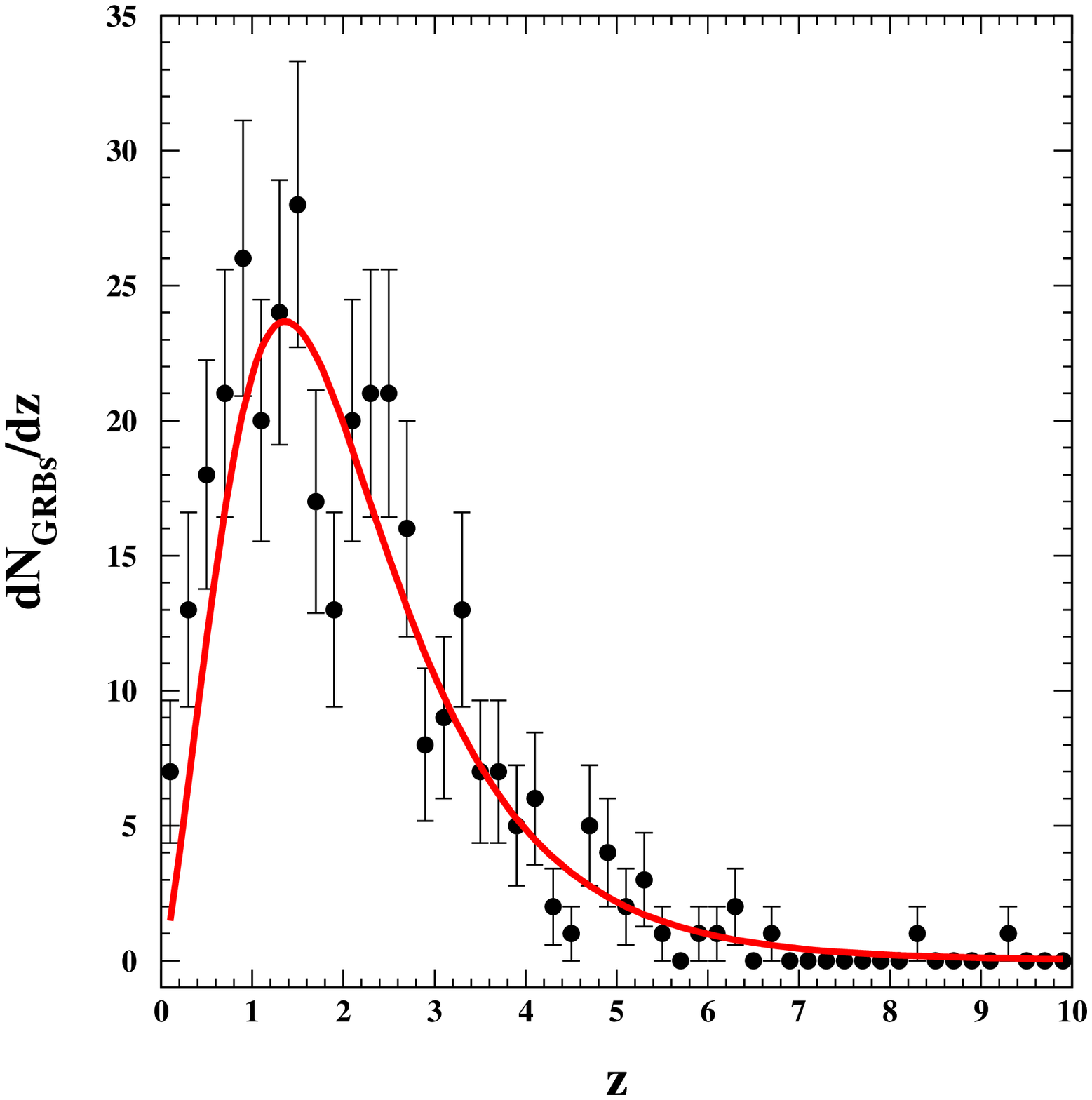,width=10.0cm,height=10.0cm}
\caption{Comparison between the redshift distribution
of 356 long GRBs with known redshift observed before June 2018
and their expected distribution in the CB model if the production rate of 
GRBs is proportional to the SFR; ($\chi^2/dof\!=\!37.57/49\!=\!0.77$).}
\label{Fig22}
\end{figure}  
{\bf In the FB models} where $\theta_j\!>>\!1/\gamma(0)$ the observed 
rate of GRBs is expected to be proportional to the star formation 
rate (SFR) [72] back to very large redshifts beyond those accessible to optical measurements.
However, the observed rates of LGRBs and XRFs do not follow the SFR. 
Unlike the SFR (in a comoving unit volume), which first increases with redshift, [73] 
the observed rate of LGRBs 
first decreases with increasing redshift in the range $z\!\leq\!0.1$, 
even after correcting for detector flux threshold [74]. 
At larger redshifts, it increases faster than the  SFR [75].
The discrepancy at small $z$ was interpreted as evidence that ordinary LGRBs 
and low-luminosity LGRBs and XRFs  with much lower  luminosity belong to 
physically distinct classes [76]. The discrepancy 
at $z\!\gg\! 1$ was claimed to be due to different evolutions [77].
Figure 23, also displays  the  cumulative distribution of GRBs as function of redshift 
in the standard fireball model, with the redshit evolution of the LGRBs relative to the SFR 
assumed in [77]. As can be seen from Figure 23, the observational data does not support    
the evolution proposed in [77]. 
\begin{figure}[]
\centering
\epsfig{file=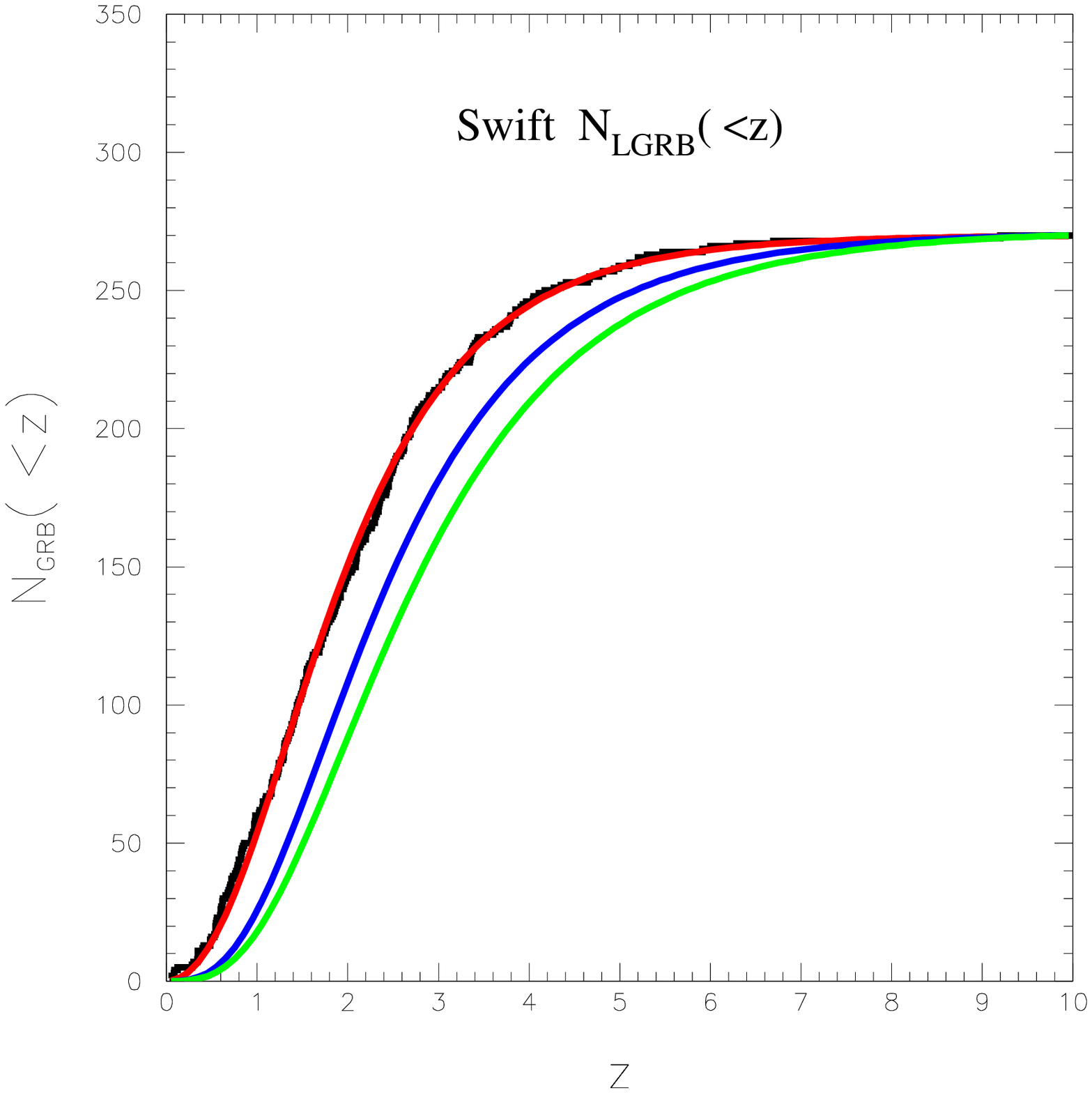,width=10.0cm,height=10.0cm}
\caption{Comparison between the cumulative distribution function, $N(\!<\! z)$,
of the 262 LGRBs with known redshift (histogram) that were detected by
Swift before 2014 and $N(\!<\!z)$ expected in the CB model (left curve)
for  long GRBs whose rate traces the SFR. Also shown is
the distributions expected in FB models with evolution (left and right 
curves) [77] and without evolution (middle curve).}
\end{figure}

\noindent
{\bf Test 11: Low Luminosity GRBs.}\\
{\bf In the CB model} the observed properties of GRBs depend strongly on 
their viewing angle relative to the CBs' direction of motion. Ordinary 
(OR) GRBs are viewed from angles $\theta \!\sim\! 1/\gamma_0$ relative to 
the CB direction of motion, which yield $\delta_0\!\sim\! \gamma$. In the 
CB model, low luminosity (LL) GRBs are ordinary GRBs with similar 
intrinsic properties, but viewed from far off-axis, i.e., 
$\gamma_0\,\delta_0\!=\!(1\!-\!\beta\cos\theta)\! 
\approx\!2/\theta^2$. Consequently, under the approximation that GRBs are
standard candles, ordinary   GRBs and LL GRBs satisfy the relations
\begin{equation}
E_{iso}(LL\,\,\, GRB)\!\approx E_{iso}(OR\,\,\, GRB)/[\gamma_0^2\,(1\!-\!cos\theta)]^3\, 
\end{equation}
\begin{equation}
L_p(LL\,\,\, GRB)\!\approx\!L_p(OR \,\,GRB)/[\gamma_0^2/(1\!-\!cos\theta)^4\,
\end{equation}.
Eqs.(22),(23) and the predicted correlations between  
properties of LL GRBs can be used to test the far 
off-axis ($[\gamma(0)\, \theta]^2\!\gg\!1$) identity  of LL-GRBs. These 
correlations include:

\noindent
a) $(1\!+\!z)E_p\propto E_{iso}^{1/3}$, which is verified in Figure 4.\\
b) $(1\!+\!z)t_b\!\propto\![(1\!+\!z)E_p\,E_{iso}]^{-1/2}$,\\  
valid for both near axis and far off-axis SN-GRBs, which is verified in 
Figure 12.\\ 
c) Production rate per comoving unit volume which is proportional to the star 
formation rate with the same proportionality constant as that for ordinary
GRBs, after correcting for viewing angle, which is verified in [73] and in 
Figure 23.\\

Perhaps, the best evidences that both low-luminosity and ordinary SN-GRBs 
belong to the same class of GRBs comes from the fact that both types of 
SN-GRBs are produced in very similar SNeIc [23] akin to SN1998bw. 
For instance, SN2013cq that produced GRB130427A at redshift $z\!=\!0.34$, 
with a record high GRB fluence measured by Swift BAT and Fermi GBM, and 
with an $E_{iso}\!\sim\!10^{54}$ erg, was very similar [23]
to SN1998bw, which produced the LL GRB980425 with a  record 
low $E_{iso}\!\sim\! 10^{48}$ erg [78], roughly smaller by a 
factor $10^6$.\\

Moreover, the best fit CB model 
light curve of the X-ray afterglow of GRB980425
measured with Beppo-SAX [79] and CXO [80] shown in Figure 24,
has yielded $\gamma\,\theta\approx 8.7$. Thus, Eq.(22) yields
$E_{iso}(GRB980425)\!\approx\! 1.84\times 10^{-5}\!\langle
E_{iso}(OR\,\,\, GRB)\rangle\!\approx\! 1.3\times 10^{48}$ erg for 
$\langle\!E_{iso}(OR\,\,\, GRB)\!\rangle \!\approx\! 7\times 10^{52}$ 
erg  in good agreement with the observed 
value [78] $E_{iso}(GRB980425)\!\approx\!(1\!\pm\!0.2)\times 10^{48}$ erg.\\

\noindent
{\bf In the FB model,} low luminosity GRBs were claimed to be intrinsically 
different from ordinary SN-GRBs and  belong to a different class [76] of 
GRBs. This has been forced upon the fireball model  because of three reasons:\\ 
a) The standard fireball model could not explain $\sim$ 6 orders of magnitudes
difference between  the value of $E_{iso}$ of low luminosity SN-GRBs,  
such as that of GRB980425, and that of very high luminosity SN-GRBs, 
such as  GRB130427A, which were produced by very similar SNeIc [23].\\    
b) The SN-GRB association and the observed locations of long GRBs in star 
formation regions of galaxies suggested [72]
that long GRBs trace the star  formation rate. However, it was found that 
the redshift distribution of LGRBs at $z\!<\!0.1$ does not follow the SFR [77]: 
while the SFR in a comoving unit volume  first increases with redshift, 
the observed rate of long GRBs in the range $z\! <\! 0.1$ decreases 
with increasing redshift.\\ 
c) The observed production rate of low luminosity GRBs at  $z\!<\!0.1$ 
relative to the SFR is  larger by a large factor than the ratio at much 
higher $z$ values [77]).

Note, however, that despite the past widespread belief 
among the FB models promoters that LL-GRBs 
and OR-GRBs belong to two distinct classes of long GRBs [76], the far 
off-axis jet origin of the low luminosity SHB170817A, has 
now been widely accepted. This followed its  measured 
$E_{iso}\!\approx\! (5.4\!\pm\!1.3) \times 10^{46}$  erg [81], which was smaller 
by roughly five orders of magnitudes than that of ordinary SHBs, and  
from the  measured viewing angle by Ligo-Virgo,
of the axis of the n*n* binary  which produced GW170817 [21], 
from the observed behavior of its late time afterglow [82], and  
from the measured superluminal motion of its point-like radio 
source  [83](see Tests 12, 13).\\

{Test 12: Superluminal Velocity in SN-GRBs.}
The first observation of an apparent  superluminal velocity 
of a source in the plane of the  sky  
was reported [84] in 1902, and since 1977 in many 
high resolution observations of highly relativistic jets 
launched by quasars, blazars, and microquasars.
The correct interpretation of such observations within the framework 
of special relativity was provided in [85].

A relativistic source with a velocity $\beta\,c$ at redshift $z$ 
which is viewed from an angle $\theta$ relative to its direction 
of motion and is timed by the local arrival times of its 
emitted photons, has an apparent velocity in the plane of the sky,
which is given by,
\begin{equation}
V_{app}\!=\!{\beta\,c\,sin\theta\over (1\!+\!z)(1\!-\!\beta\,cos\theta)}
\!=\!{\beta\,\gamma\, \delta\, c\, sin\theta \over (1\!+\!z)}.
\end{equation}
For $\gamma\!\gg\!1$, $V_{app}$ has a maximum value
$\approx\! 2\,\gamma\, c/(1\!+\!z)$   at $sin\theta\!=\! 1/\gamma$.

The predicted superluminal velocity of the jetted CBs, which produce 
GRBs cannot be verified during the prompt emission phase, mainly 
because of its short duration and the large cosmological 
distances of GRBs. However, the superluminal velocity of the jet in far 
off-axis, i.e., nearby low luminosity GRBs, can be obtained from high 
resolution follow up measurements of their afterglows [86].
Below, we discuss two cases.\\ 

\noindent
{\bf a. GRB980425.} So far, 
the radio and X-ray afterglow of GRB980425, the nearest observed 
SN-GRB with a known redshift, $z\!=\!0.0085$, has offered the 
best opportunity to look for the superluminal signature of the 
highly relativistic jets, which produce GRBs [86].
To the best of our knowledge, this has been 
totally overlooked in the 
late-time high resolution X-ray [80]  and 
radio observations [87] of
SN1998bw/GRB980425 due to biases. But if these transient sources observed
on day 1281 and 2049.19, respectively, in the same direction from 
SN1998bw, are the CB which produced GRB980425, then it moved
with an apparent superluminal jet velocity of $V_{app} 
\!\approx\! 340\, c$ and a viewing angle 
$\theta\!\approx\!2\,c/V_\perp\!\approx\!1/170$ rad [85].
This interpretation implies  that 
these sources are not present there anymore, and were 
not there before SN1998bw/GRB980425.

Supportive evidence for this CB model value of  $V_{app}$ 
in GRB980425 is provided by other observations.\\
(i) The expected value $E_p\!\approx\!\epsilon_p\,\gamma\,\delta/
(1\!+\!z)\!\approx\!1\, eV/1.0085\,(1\!-\!cos\theta)\!\approx 57$ keV, 
which is in good agreement with the observed [78] value 
$E_p\!=\!55\!\pm\!21$ keV.\\
(ii) The ratio of the observed $FWHM\!\approx\! 
R_g\theta^2/c\!\approx\!12$ s  duration of GRB980425 
and the observed mean $FWHM\!\approx\!\langle 2\,(1\!+\!z)\
R_g/\gamma^2\,c \rangle \!=\!0.89$ s  duration of 
ordinary GRBs [87] yield  $\gamma\,\theta\!\approx\!9$ for 
the observed average $\langle\!1\!+\!z \rangle\!\approx\!3$. 
Thus, Eq.(21) yields  $E_{iso}(GRB980425)\!\approx\!1.1\times 
10^{48}$ erg, in good agreement with its measured value [78] 
$E_{iso}(GRB980425)\!\approx\!(1.0\!\pm\!0.2) \times 10^{48}$ erg.\\
(iii) The 0.3-10 keV X-ray light-curve of the afterglow of 
GRB980425 measured by Beppo-SAX [79] and CXO [80]  can be  
well fit by the CB model with $\gamma\, \theta\!\approx\! 8.7$ 
as shown in Figure 24.
\begin{figure}[]
\centering
\epsfig{file=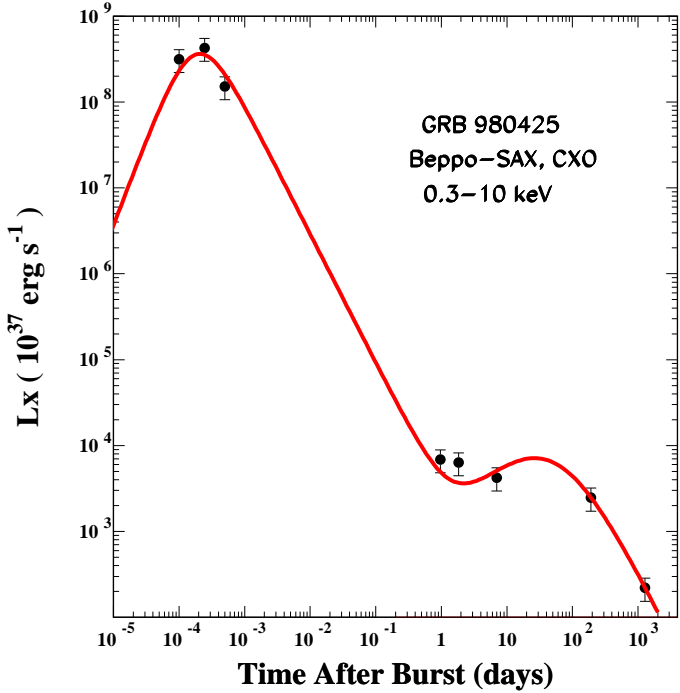,width=10.0cm,height=10.0cm}
\caption{The 0.3-10 keV X-ray light-curve of GRB980425 measured by Beppo-SAX 
[79] (first 7 points). The last point at 1281 days is 
due to the source S1b resolved by CXO [80].
The line is the CB model best fit light-curve to the prompt emission 
pulse and the  afterglow of GRB980425 for $\gamma\,\theta\!\approx\! 8.7$.}
\end{figure}  

\noindent
{\bf b.  GRB030329.} The relative proximity 
($z$=0.1685) of GRB030329 and its record-bright radio afterglow 
made possible its record long, high resolution follow-up 
observations with the  Very Long Baseline Array (VLBA) and  
Very Long Baseline Interferometry (VLBI), until 
3018.2 days post GRB [88]. Assuming a disk shape, the radio image 
of GRB030329/SN2003dh was fit with a circular Gaussian of 
diameter $2\,R_\perp(t)$. The mean apparent image expansion,
before time $t$, is 
\begin{equation} 
\langle \beta_{app}\rangle\!=\!2\,R_\perp/c\,t\,. 
\label{Eq25} 
\end{equation} 
SN2003dh and GRB030329, however, had individual image sizes much 
smaller than the resolution of the VLBA and VLBI arrays: The 
initial large expansion velocity of broad-line SNeIc, such as 
SN2003dh decreases to less than 10,000 km/s within the first 
month, beyond which it continues to decrease, roughly like 
$t^{-1/2}$. Such an expansion velocity yields SN image-sizes 
$<0.002$ pc and $<\!0.005$ pc on days 25 and 83 after burst, 
compared to the joint image-size of GRB030329/SN2003dh, $\sim$0.2 
pc and $\sim$ 0.5 pc, respectively, extracted from the radio 
observations [88]. As for CBs, time dilation and ram pressure 
suppress their lateral expansion in the circumburst medium, as 
long as they move with a highly relativistic speed. Although 
their small size implies radio scintillations, the large 
time-aberration wash them out by the time-integrated 
measurements in the observer frame -- the typical $dt\!\sim\! 100$ 
minutes integration time of the VLBA observations [88] 
corresponds to an early effective image size  $V_{perp}\,dt\!
\geq\! 10^{16}$ cm.

The VLBA and VLBI measurements [89] could not resolve the 
separate  images of SN2003dh and the CB, which produced GRB030329 
and its afterglow, nor the  superluminal displacement  the CB 
away from SN2003dh. If, however, the size of their joint radio 
image and its expansion rate as measured in [89] 
are adapted as a rough estimate of the time-dependent distance 
between the GRB afterglow and the SN, they can be used to test 
the CB model, as shown below.

In ordinary GRBs, CB deceleration in an ISM with a constant density 
yields the late-time behaviors $\delta(t)\!=\!2\,\gamma(t)\propto t^{-1/4}$.
Consequently, $F_\nu\propto t^{-\alpha_\nu}\,\nu^{-\beta_\nu}$ with 
$\alpha_\nu\!=\!\beta_\nu\!+\!1/2$, and $V_{app}\!=\! 2\,\gamma^2\,\theta\, 
c /(1\!+\!z)\propto t^{-1/2}$, which are all well satisfied 
by the  late-time X-ray afterglow [90] and radio observations of 
of GRB030329 [89]. E.g., the measured spectral index 
$\beta_X\!=\!1.17\!\pm\! 0.04$ 
in the 0.2-10 keV X-ray band [88], yields a late-time temporal decay index 
$\alpha_x=\beta_x\!+\!1/2\!=\!1.67\!\pm\!0.04$, in good agreement with 
the observed $\alpha_x\!=\!1.67$ [90] 
as shown in Figure 25. 
\begin{figure}[]
\centering
\epsfig{file=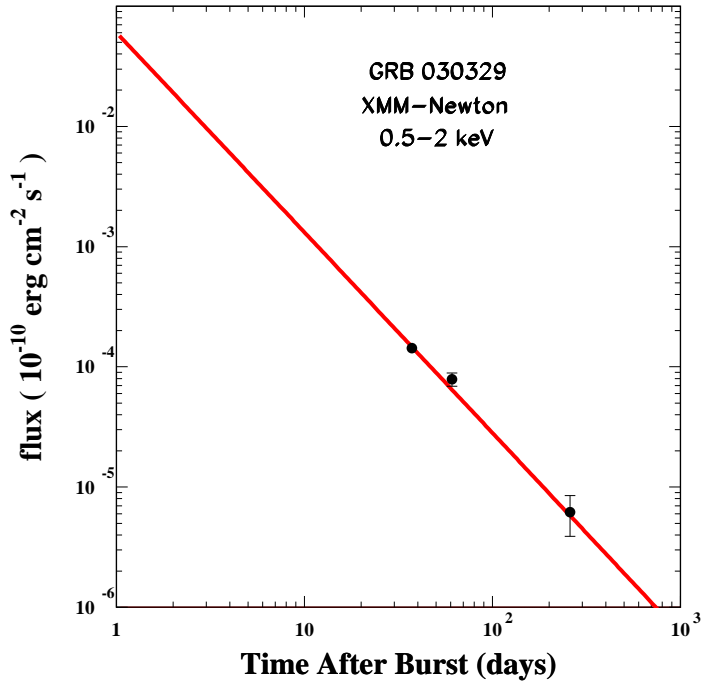,width=10.cm,height=10.0cm}
\caption{The late-time 0.5-2 keV X-ray afterglow of the joint source 
GRB030329/SN2003dh as measured by XMM-Newton [91]. The line is the CB 
model prediction for the X-ray light-curve assuming that 
$\alpha_x=\beta_x+1/2$  where $\beta_x=1.17\pm 0.04$.}
\end{figure}
The late time VLBA and VLBI radio measurements of the image-size of 
GRB030329/SN2003dh are also in good agreement with the CB model 
prediction.
\begin{equation} 
\langle V_{app}(<t)\rangle\! \approx\! 2\, V_{app}(t)\!\propto\! t^{-1/2}\, 
\end{equation}
This is shown in Figure 26. The CB model prediction is for $\gamma_0\, 
\theta=1.76$ obtained from Eq.(22) for the observed [89] 
$E_{iso}(GRB030329)\!\approx\! (1.86\!\pm\! 0.08)\times 10^{52}$ and 
$\langle E_{iso}(OR\,\,\, GRB)\rangle \!\approx\! 7\times 10^{52}$ erg 
assuming a standard candle GRBs with 
$\epsilon_p\!\approx\!1$ eV. The late-time universal behavior, 
$V_{app}\!\propto\! t^{-1/2}$, is valid as  
long as the CB moves within a constant density, independent of 
the specific values  of the density and  of $\epsilon_p$. The 
observed late-time behavior shown in Figure 26, suggest  
deceleration in edge-on host galaxy of a CB with 
$\gamma(0)\!\approx 400$.  
\begin{figure}[]
\centering
\epsfig{file=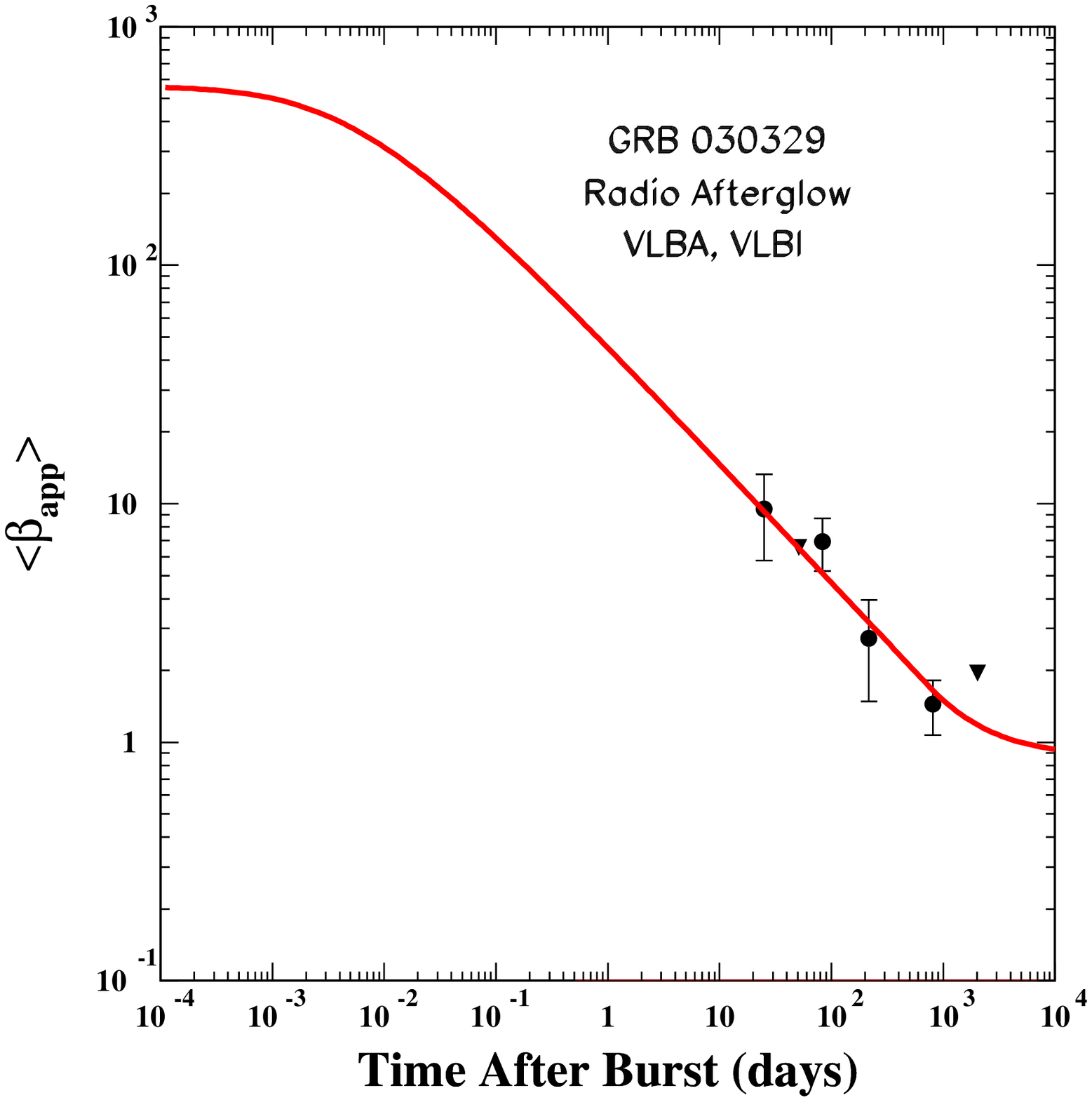,width=10.0cm,height=10.0cm}
\caption{The time-averaged expansion rate of the radio image of 
GRB030329/SN2003dh [88]. The line is the predicted 
$\langle \beta_{app}\rangle$ of the CB, which produced GRB030329 
and its afterglow, assuming that $2R_\perp$ in Eq.(25) is its distance  
from SN2003dh.}
\end{figure}

\noindent
{\bf The FB model} has been used to provide  different posteriori
interpretations of the observed superluminal expansion [88] of the image 
size of the source of the radio afterglow of GRB030329/SN2003dh, as 
the observations progressed. All of them were parametrizations, which
depend on many free adjustable parameter rather than falsifiable 
predictions. Moreover, the radio VLBI observations of SHB170817A
show a pointlike (unresolved) radio source displaced at a superluminal 
speed from the SHB location, rather than an increasing size of a radio 
image around the location of the SHB as predicted  by spherical fireball 
models  or conical fireball models (i.e., structured conical jets) 
misaligned with respect to the axis of the n*n* binary whose merger 
produced SHB170817A.     

\section{GRB Theories confront SHB170817A}
GW170817 was the first binary neutron-star merger detected with 
Ligo-Virgo [21] in gravitational waves (GWs). It was followed by 
SHB170817A, $1.74\!\pm\! 5$ s [20] after the end of the GWs detection, 
with an afterglow across the electromagnetic spectrum, which was used to 
localize it [92] to the galaxy NGC 4993 at a distance of $~40$ Mpc. 
The GW170817/SHB170817A association was the first indisputeable 
confirmation that n*n* mergers in compact binaries due to GW emission 
produce GRBs. That was first suggested in 1984 [5] to be due to 
explosion of the lighter n* after tidal mass loss (later called Macornova 
[6]), and in 1987 due to neutrino annihilation fireball [7] around the 
remnant n*. In 1994, the fireball mechanism was replaced [11] by 
ICS of external light by a highly relativistic jet of ordinary plasma 
launched by fall back ejecta on the remnant star after the merger, which 
became  the basis of the  CB model of SGRBs and their afterglows.

The relative proximity of SHB170817A provided many critical tests of
SGRB theories. Two days  before the GW170817/SHB170817A 
event, the CB model of GRBs was used to
predict [93] that most of 
the SGRBs associated with Ligo-Virgo detections of n*n* mergers will 
be beamed far off-axis. This is because of beaming and the relatively small 
volume of the Universe from where n*n* mergers could be detected by 
the current Ligo-Virgo detectors. 
Consequently, only a small fraction of 
them would be visible as low luminosity far off-axis SGRBs [93], 
while the early time universal afterglow powered by the spin down  
of the newly born MSP will produce a characteristic isotropic 
afterglow with a universal light-curve [22], visible from all 
SHBs associated with Ligo-Virgo detections of n*n* mergers.

Shortly after the detection of the low-luminosity SHB170817A, several 
authors claimed [81] that SHB170817A was an ordinary near axis SGRB 
despite its observed smaller $E_{iso}$ by four orders of magnitude 
than that of typical SGRBs. Moreover, many authors and promoters of 
fireball and firecone models, who claimed in the past that ordinary 
and low luminosity GRBs belong to two different classes of GRB, 
rushed to suggest the opposite for the low luminosity SHB170817, 
namely, that it was viewed from far off-axis. This change was not 
enough, and drastic changes in the FB models were introduced under 
the cover of "structured jets". Despite the use of many free 
adjustable parameters, all such models did not predict, rather than 
posdict, correctly the late-time behaviour of the light curve of the 
afterglow of SHB170817A.

\noindent
{\bf Test 13: SHB170817A Properties}\\ 
{\bf Jet Geometry.} The VLBI/VLBA radio observations of the late time 
radio afterglow [82] of SHB170817A provided the first successful 
measurement of the the image of the highly relativistic afterglow source 
of a GRB separated from the GRB and escapes from it with an apparent 
superluminal speed, as predicted by the CB model already two decades ago 
[28] in analogy with blazars and microquasar ejections. This is in 
contrast to the unresolved expanding image of GRB and afterglow [89] 
advocated by the fireball and firecone models of GRBs and GRB afterglow. 
This is shown in Figure 27 adapted from [82]. It demonstrates a 
displacement with time of a point-like source as assumed in the CB model 
(and seen before in microquasars and blazars [28]), rather than an increasing 
unresolved joint image of the GRB and its afterglow expected in the FB 
models and claimed before in the case of GRB030329/SN2003dh [89]. As 
shown in Figure 27, the angular location of the source of the radio afterglow 
of SHB170817A has moved in the plane of the sky during $\Delta t$=155 d 
(between day 75 and day 230) by  $\Delta\theta_s\!=\!2.68\!\pm\!0.3$ mas [82].

\begin{figure}[]
\centering
\epsfig{file=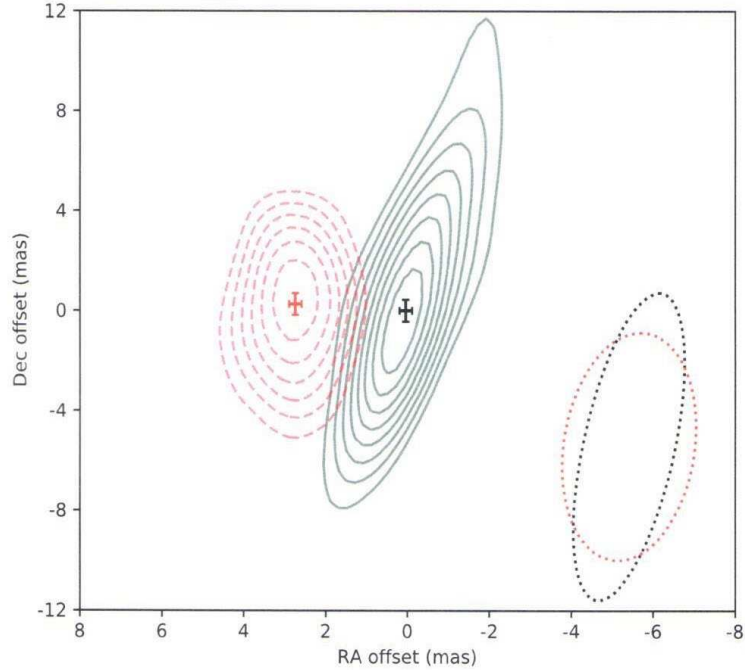,width=10.cm,height=10.cm}
\caption{Proper motion of the radio counterpart of GW170817. The 
centroid offset positions (shown by $1\sigma$ errorbars) and 
$3\sigma$-$12\sigma$ contours of the radio source detected 75 d 
(black) and 230 d (red) post-merger with Very Long Baseline 
Interferometry (VLBI) at 4.5 GHz. The radio source is consistent with 
being unresolved at both epochs. The shape of the synthesized beam 
for the images from both epochs are shown as dotted ellipses to the 
lower right corner. The proper motion vector of the radio source has 
a magnitude of $2.7\!\pm\!0.3$ mas and a position angle of 
$86\!\pm\!18$ deg over 155 d. The Figure was adapted from [82].}
\end{figure}

\noindent
{\bf Viewing angle from superluminal motion}\\
Assuming a highly relativistic ($\beta\!\approx\! 1$) point like radio 
source, its superluminal velocity satisfies  
\begin{equation} 
V_{app}\!\approx\! 
{c\,sin\theta\over (1\!+\!z)\,(1\!-\!cos\theta)}\!\approx\! 
{D_A\,\Delta \theta_s\over (1\!+\!z)\Delta t}\,. 
\end{equation} 
For an angular distance $D_A\!=\!39.6$ Mpc to SHB170817A in NGC 4993 at 
a redshift $z\!=\! 0.009783$ [92], which follows from the local value 
of the Hubble constant $H_0\!=\!73.4\!\pm\! 1.62\,{\rm km/s\, Mpc}$ 
obtained from Type Ia supernovae [94], and the value $25\pm 4$ deg, 
which was obtained [95] from GW170817 and its electromagnetic location 
[81,92], Eq.(27) yields $V_{app}\!\approx\! (4.0\!\pm\! 0.40)\,c$ and 
consequently $\theta\!\approx\! 28\!\pm\! 2$ deg . This value of the 
viewing angle $\theta$ is in agreement with the value $\theta\!=\!25\pm 
4$ deg, obtained [95] from GW170817 and its electromagnetic location 
[81,92], assuming the local value of $H_0$ [94] and that the CB 
was ejected along the rotation axis of the n*n* binary.

\noindent
{\bf Initial Lorentz factor}\\
{\bf In the CB model,} SGRBs like LGRBs are assumed to be standard 
candles viewed from different angles and therefore satisfy similar 
correlations. In particular, in the CB model, low luminosity (LL) 
SHBs such as SHB170817A are ordinary SHBs that are viewed from far 
off-axis (FOA). Consequently, their $E_{iso}$ is expected to 
satisfy,
\begin{equation}
E_{iso}(FOA)\! \approx\! \langle\! E_{iso}(OR)\!\rangle\,/
[\gamma^2\,(1\!-\!cos\theta)]^3 \,,
\end{equation}
while
\begin{equation}
(1+z)\,E_p(FOA)\!\approx\! \langle (1+z)\,E_p(OR)\rangle\,/
[\gamma^2\,(1\!-\!cos\theta)]\,.
\end{equation} 
The measured value  $E_{iso}\!\approx\! 5.4\times 10^{46}$ erg of 
SHB170817A [96], the mean 
value $\langle E_{iso}\rangle\!\approx\!1\times 10^{51}$ erg of 
ordinary (OR) SGRBs,  and  the  viewing angle 
$\theta\!\approx\!28$ deg obtained  from the observed superluminal 
velocity of the source of its radio afterglow [82], Eq.(29)  
yields  $\gamma_0\!\approx\!14.7$ and $\gamma_0\theta\!\approx\! 7.2$.\\

\noindent
{\bf Radio source size}\\
{\bf In the CB model}, the initial expansion velocity of a CB  
in its rest frame is $c/\sqrt{3}$, the speed of sound in a relativistic gas.
In the observer frame   
\begin{equation}
V_{exp}(t)\!\leq\!\delta\,c/(1+\!z\!)\sqrt{3}\!\approx\! c/(1\!+\!z\!)
\sqrt{3}\, \gamma\,(1\!-\!cos\theta)
\end{equation}
Consequently, in the CB model, the radius of the CB on day 230 
after burst is much  below $2\times 10^{17}$ cm.
At the angular distance of $D_A\!=\!39.6$ Mpc of NGC 4993,
it has an angular size much below the VLBI 
upper limit $\approx\!1 mas$ ($6\times 10^{17}$ cm) on the size of the radio 
source  at the distance of NGC 4993 on day 230.

\noindent
{\bf In the FB models}, the opening angle of the jet has been adjusted 
to obey the VLBI constraint, but the consistency of the adopted jet 
structure and the VLBI image has not and could not been demonstrated. 
 
\noindent
{\bf Prompt Emission Correlations}\\
{\bf In The CB model,} the canonical CB model correlations
as given by Eqs.(3),(4) are satisfied well separately by each of 
the 3 major types of 
GRBs; SN-LGRBs, SN-less LGRBs and SN-less SGRBs, as was demonstrated 
in Figures 3,4,5. Eqs.(28),(29) together with the viewing angle of 
the highly relativistic CB obtained from its apparent superluminal 
motion [82], allow additional tests of CB model predictions,
as follows.\\
\noindent 
{\bf The peak energy of SHB170817A}\\
Assuming the same redshift distribution of GRBs and SHBs
with a mean value $z\!\approx\!2$, and the observed mean value
$\langle\! E_p\!\rangle\!=\!650$ keV in SHBs [96],
yields  $\langle(1\!+\!z)\,E_p\rangle\! \approx\! 1950$
keV. Consequently, Eq.(29) with $\gamma_0\theta\!\approx\! 7.2$
yields  $(1\!+\!z)\,E_p\!\approx\!75$ keV for SHB170817A,
compared to $(1\!+\!z)E_p\!=\!82\!\pm\!23$
keV ($T_{90}$) reported in [20a],  
$185\!\pm\!65$ keV estimated in [20c],
and  $E_p\!\approx\!
65\!+\!35/\!-\!14$ keV estimated in [20d], 
from the same data, with a mean value
$(1\!+\!z)E_p\!=\!86\!\pm\!19$ keV.

In the CB model the peak time $\Delta t$ after the beginning 
of a GRB/SHB pulse is roughly equal to half of its FWHM.
Assuming that SHBs are rougly standard candles, the
dependence of $\Delta t$ on viewing angle of the CB direction of 
motion is given by 
\begin{equation}
\Delta t(LL\,\,\,SHB)\,\approx\! \gamma_0^2\,(1\!-\!cos\theta)
\langle \Delta t(SHB)\rangle,
\end{equation}
where $(SHB)$ stands for OR-SHB. For a viewing angle  $\theta\!\approx\!28$
deg, obtained from the superluminal motion of the resolved source of 
the late-time radio afterglow of SHB170817A,     
$\Delta t\!\approx\! 0.58$ s obtained  from the prompt emission pulse 
of  SHB170817A (see Figure 9), and $\langle FWHM(SHB)\rangle$=55 ms, 
Eq.(28) yields $\gamma_0\!\approx\! 14.7 $. This value is  consistent 
with the values obtained from either the estimated
$E_{iso}\!=\!5.4\times 10^{46}$ of SHB170817A in [81], 
or the weighted mean value $E_p\!=\! 86\!\pm\! 19$ keV from its 
estimates in [22].

Moreover, in the CB model the shape of resolved SHB and GRB  pulses 
satisfies $2\,\Delta t\!\approx\! FWHM \!\propto\! 1/E_p$. Using the observed 
$\langle FWHM(SHB)\rangle\!\approx\!55$ ms and $\gamma_0\!=\!14$, in OR 
SHBs, and  $\theta\!\approx\! 28$ deg, Eq.(31) yields $\Delta t 
(SHB170817A)\!\approx\!0.63$ s, in good agreement with its observed value, 
$0.58\!\pm\! 0.06$ s. 

We caution, however, that replacement of physical 
parameters by their mean values may be only indicative but not completely reliable, 
because of detection thresholds, selection effects and wide 
spread values of physical parameters.

Perhaps the simplest correlation expected in the CB model to be satisfied 
by resolved pulses of SGRBs is  
\begin{equation} 
E_p\!\propto\! 1/T_p 
\end{equation} 
where $T_p$ is the duration of the pulse (or its peak-time $\Delta t$).  This 
correlation, which is expected to be satisfied also by resolved LGRB pulses, is 
easier to test in SGRBs because, unlike LGRBs, a large fraction of SGRBs are 
single pulse SHBs or a sum of very few resolved pulses. Figure 28  compares the 
predicted correlation  $E_p\, T_p\!\approx\! 100\,keV\,s $ and  that obtained 
from the reported  values of $E_p$ and $T_p$ in the GCN circulars archives,
for resolved SGRB pulses measured by the Konus-Wind and by the Fermi-GBM collaborations. 
As indicated in Figure 28, this CB model correlation seems to be  satisfied
by most of the measurements, in particular by those with small 
estimated observational errors. 
\begin{figure}[]
\centering
\epsfig{file=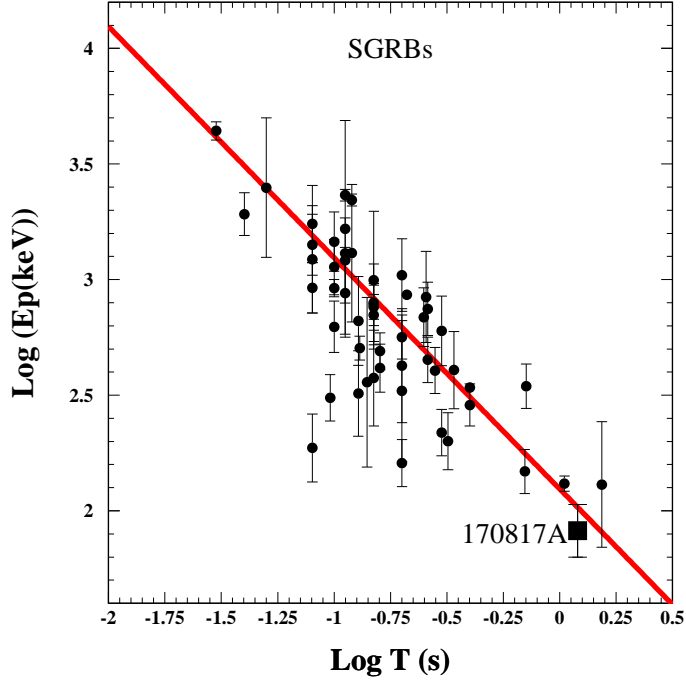,width=10.0cm,height=10.0cm}
\caption{Comparison between the predicted  correlation  $E_p\, 
\approx\! 100\,keV/(T/s) $ and that obtained from the 
reported  values of $E_p$ and $T_p$ in the GCN circular archive
for resolved pulses in 54 SGRB pulses measured by the Konus-Wind 
and Fermi-GBM collaborations.}
\end{figure}  

\noindent
{\bf The Early Time Afterglow.}\\
The bolometric afterglow of SHB170817A, 
during the first few days after burst, has the universal shape of 
the early time X-ray afterglow of all SGRBs and SN-less LGRBs
with well sampled X-ray 
afterglows during the first few days after the prompt emission. 
This universal shape is that expected from a PWN emission powered 
by the rotational energy loss of the newly born MSPs in NSMs, 
through magnetic dipole radiation (MDR), relativistic winds and 
high energy particles. This was shown For SGRBs in Figure 21.
In Figure 29, it is shown separately for the bolometric light 
curve [71] of SHB170817A during the first two weeks after burst.
\begin{figure}[] 
\centering 
\epsfig{file=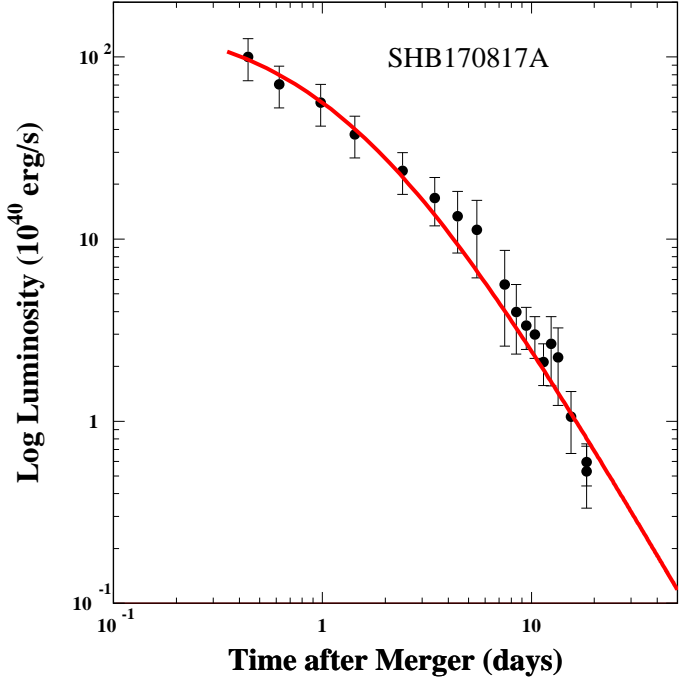,width=10.0cm,height=10.0cm} 
\caption{Comparison between the observed [71] bolometric light curve 
of SHB170817A and the universal light curve as given by Eq.(21)
assuming MSP  with $L(0)\!=\!2.27\times 10^{42}$ erg/s and,  
$t_b\!=\!1.15$ d, with an entirely satisfactory $\chi^2/dof\!=\!1.04$.} 
\end{figure}

\noindent
{\bf The Late-time Afterglow of SHB170817A.}\\
{In the CB model}, as long as $[\gamma(t)]^2\!\gg\! 1$,
and consequently $\gamma\delta(t)\!\approx\!1/(1\!-\!cos\theta)$,
the spectral energy density of the
unabsorbed synchrotron afterglow produced by a CB, which is
given by Eq.(8),  can be rewritten as 
\begin{equation}
F_{\nu}\propto n^{\beta_\nu+1/2}\,[\gamma(t)]^{1-\beta_\nu/2}\,\nu^{-\beta_\nu}\,.
\label{Eq33}
\end{equation} 
In Eq.(33),  $n$ is the baryon density of the external medium encountered
by the CB at a time $t$, and  $\beta_\nu$ is the spectral index of the       
emitted synchrotron radiation.
For a constant density, the deceleration of the CB yields a late-time
$\gamma(t)\propto t^{-1/4}$ [50], and as long as $\gamma^2\!\gg\!1$, 
\begin{equation}
F_\nu\!\propto t^{1-\beta_\nu/2}\approx t^{0.73 \pm 0.03}\,, 
\end{equation}  
where we have used the observed value [83] $\beta_\nu\!=\!0.54\!\pm\!0.06$, 
which extends from the radio (R) through the optical (O) to the X-ray band.
When the CB exits the disk of NGC 4993 nearly perpendicular to it, into the 
halo of the host galaxy,
as seems to be the case in SHB170817A [92], the CB deceleration rate diminishes 
and  $\gamma(t)$ becomes practically constant. Hence, the dependence of $F_\nu$ on 
the density  turns its increase like $t^{1-\beta_\nu/2}$  
into a fast decay $\propto\![n(r)]^{(1\!+\!\beta_\nu)/2}$. Approximating the 
disk density as a function of distance $h$ perpendicular to the disk by 
$n(h)\!=\!n(0)/(1\!+\!exp(h/d))$, where $d$ is the "skin depth" of the disk, 
the light curve of the  afterglow of  SHB170817A can be approximated by,
\begin{equation}
F_\nu(t)\!\propto\!{(t/t_e)^{1\!-\!\beta_\nu/2}\,\nu^{-\beta_\nu}\over
[1+exp[(t\!-\!t_e)/w]]^{(1\!+\!\beta_\nu)/2}}
\end{equation}
where $t_e$ is roughly the escape time of the CB from the   
galactic disk into the halo after its launch. The predicted behvior 
as given by Eq.(35) is compared to the observed late-time X-ray [97] and
radio [98] afterglows of SHB170817A in Figures [30] and [31],  
respectively. 
\begin{figure}[]
\centering
\epsfig{file=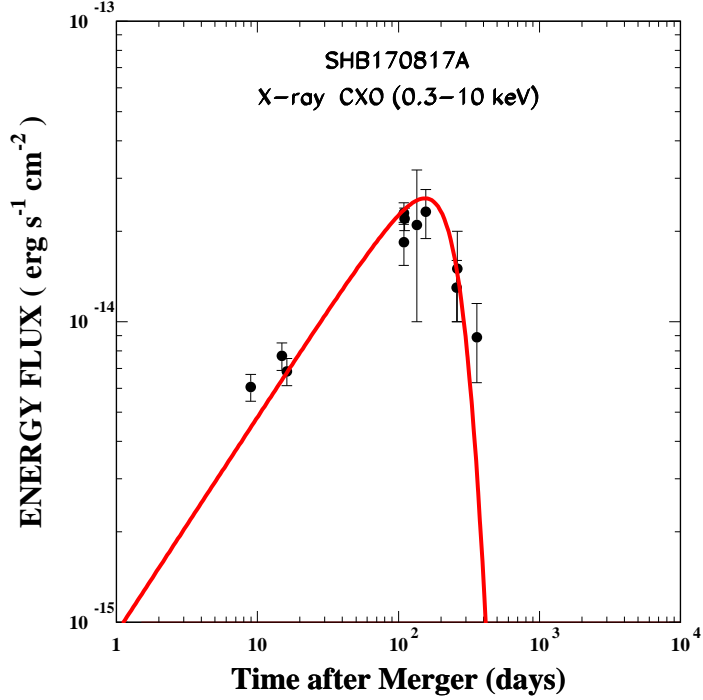,width=10.cm,height=10.cm}
\caption{The light curve of the X-ray 
afterglow of SHB170817A  measured [97] with the CXO 
and the light curve predicted by eq.(35)  for $\beta_X=0.56$ ,
$t_e\!=\!245.6$ d  and $w=63.4$ d.}
\end{figure}
\begin{figure}[]
\centering
\epsfig{file=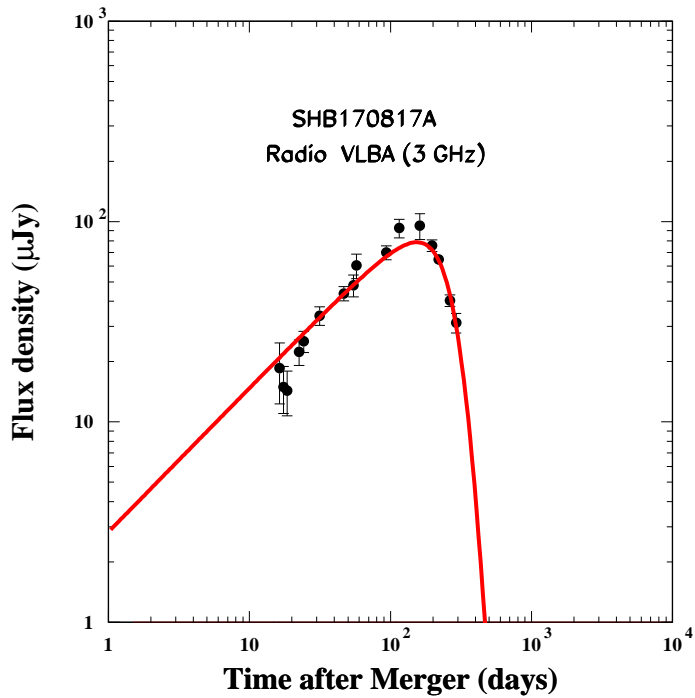,width=10.cm,height=10.cm}
\caption{The measured [98] light curve of the radio afterglow of 
SHB170817A at 3 GHz compared to the light curve predicted by eq.(35) 
with $\beta_r=0.56$, $t_e\!=\!245.6$ d  and $w\!=\!63.4$ d.}
\end{figure}

\noindent
As a further test of the predicted late-time ROX afterglow of SHB170817A 
by the CB model, we have compared Eq.(35) to the 
late-time observations of its optical afterglow with the Hubble 
Space Telescope, which extended until one year after burst [99]. 
This is shown in Figure 32.\\
\begin{figure}[]
\centering
\epsfig{file=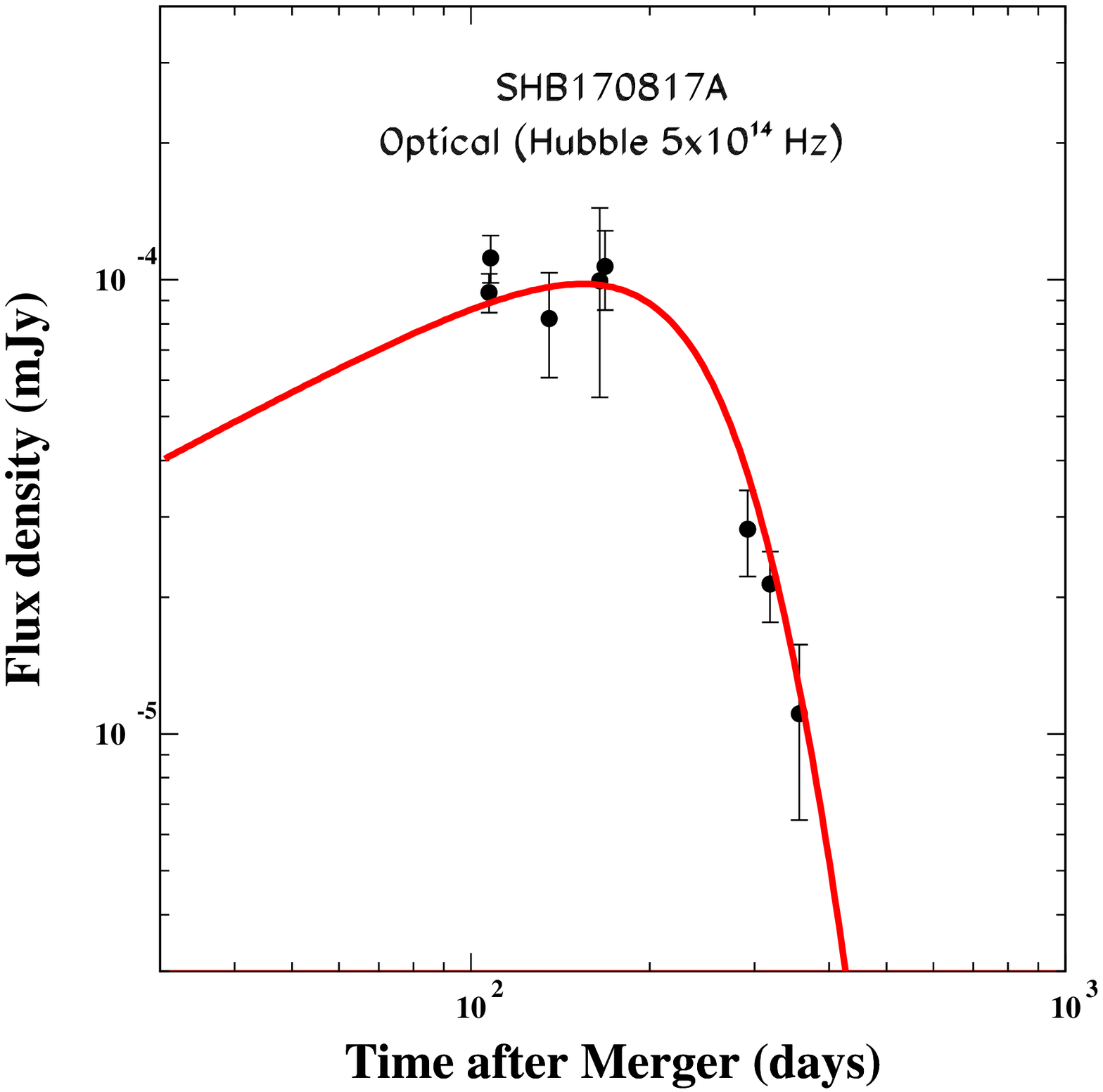,width=10.cm,height=10.cm}
\caption{The measured late-time
light curve of the optical afterglow of SHB170817A 
with the Hubble  space telescope [99] at $3.8\times10^{14}$ Hz 
and $5.1\times 10^{14}$ Hz 
compared to the light curve predicted by eq.(35) with $\beta=0.56$,
$t_e\!=\!245.6$ d  and $w\!=\!63.4$ d. ($\chi^2/dof\!=\!0.88$).}
\end{figure}

\noindent
{\bf FB Model Interpretations}\\  
Soon after the discovery of the rising late-time radio, optical and 
X-ray afterglows of SHB170817A, many versions of FB models and 
postdicted rising light curves began to flood the arXiv and the leading 
astrophysical journals. In fact, all the predicted 
FB model light curves turned out later to be failures, despite 
the fact that they involved many parameters and free choices,
which were adjusted by best fit to the available data at the time of 
publication. Below, we shall demonstrate a couple of such repeated 
failed efforts by astrophysicists and observers, promoters of the 
FB model.
  
Consider first Figure 33 (Figure 1 adapted from [100]). It 
summarizes a consensus posted in November 2017  by 25 authors 
who include main leading observers and astrphysicists, 
promoters of the FB models of GRBs, who have 
concluded that "the off axis jet scenario as a viable explanation of 
the radio afterglow of SHB170817A is ruled out" and  a chocked 
jet cocoon model is most likely the origin of gamma rays and 
afterglow of SHB170817A.
\begin{figure}[] 
\centering 
\epsfig{file=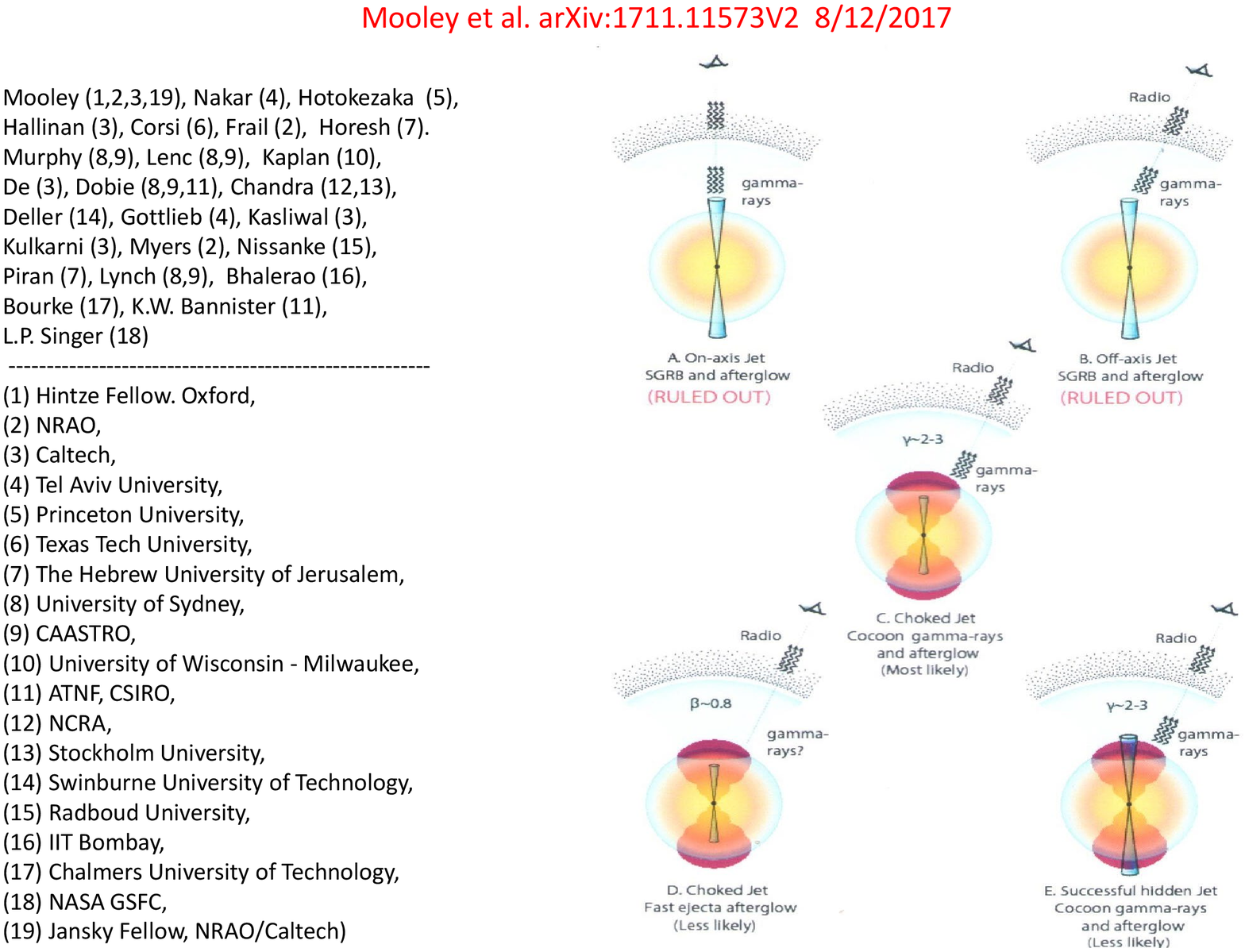,width=10.cm,height=10.cm} 
\caption{Figure 2 adapted from [100] is a cartoon representing the 
consensus of 25 authors, who include many leading observers and many other 
promoters of the FB models of GRBs who posted it in November 2017. 
They concluded  that "the off axis jet scenario as a viable explanation 
of the radio afterglow of SHB170817A is ruled out" while "a chocked 
jet cocoon model is most 
likely the origin of gamma rays and afterglow of SHB170817A".}
\end{figure}
Figure 34 (Figure 3 adapted from [100]) presents the "observational 
support" there for their claim that "the off axis jet scenario as a 
viable explanation of the radio afterglow of SHB170817A is ruled out" 
on the basis of their shown best fits to the 3 GHz radio data 
obtained before November 2017. 
\begin{figure}[]
\centering
\epsfig{file=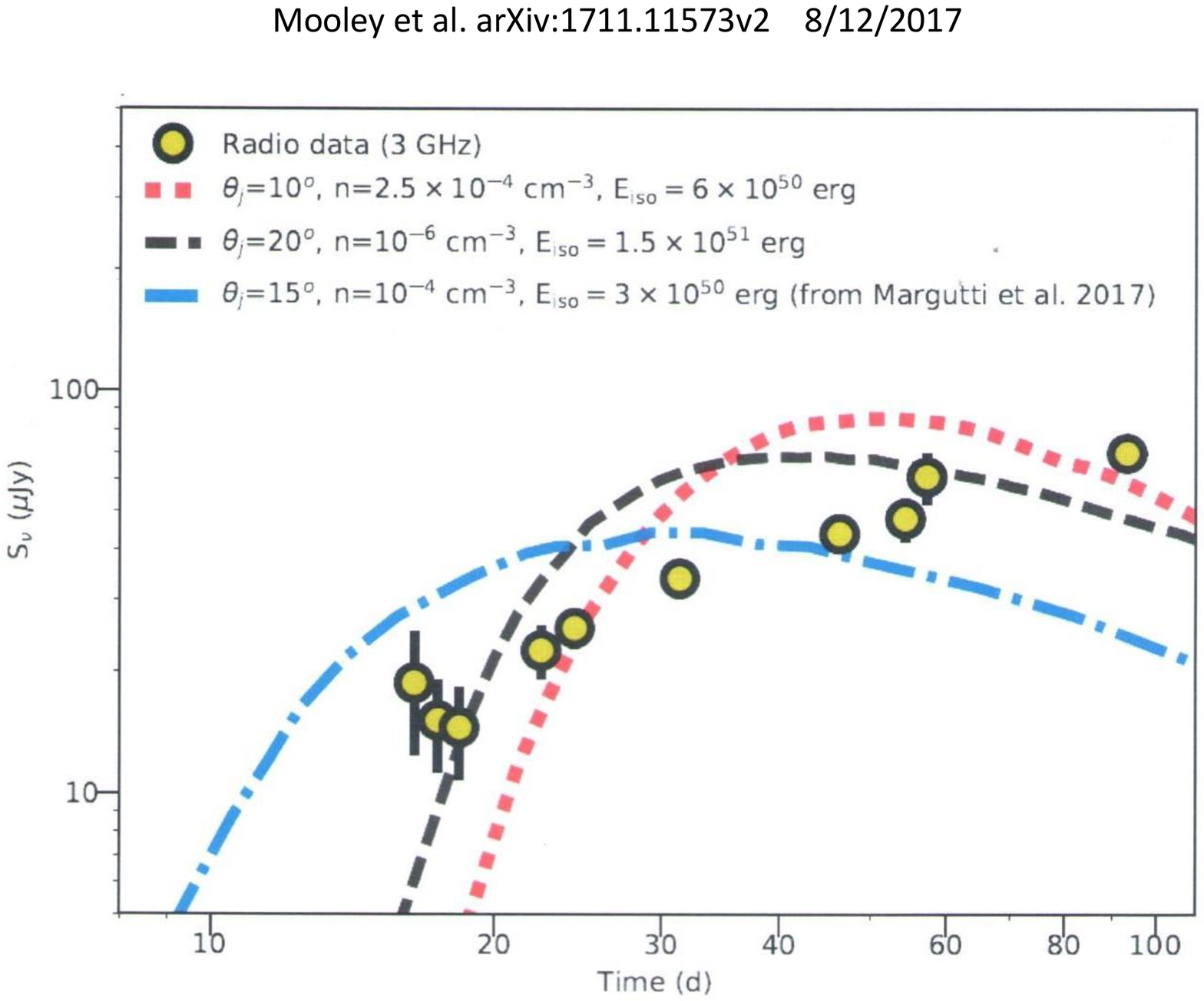,width=10.cm,height=10.cm}
\caption{Best fit light curves to the observed 3 GHz radio 
afterglow of SHB170817A measured until the end of October 2017 
[100], based on  far off-axis jet models. These fits  were used in 
[100] to conclude that "the off axis jet scenario as a viable 
explanation of the radio afterglow of SHB170817A is ruled out".}
\end{figure}
Figure 35 (Figure 1 adapted from [101]) presents a best fit smoothed 
broken power-law to the light curve of SHB170817A. The data is  from 
ATCA (circles) and VLA (squares) observations.
Unlike [100].  
\begin{figure}
\centering
\vspace{2cm}
\epsfig{file=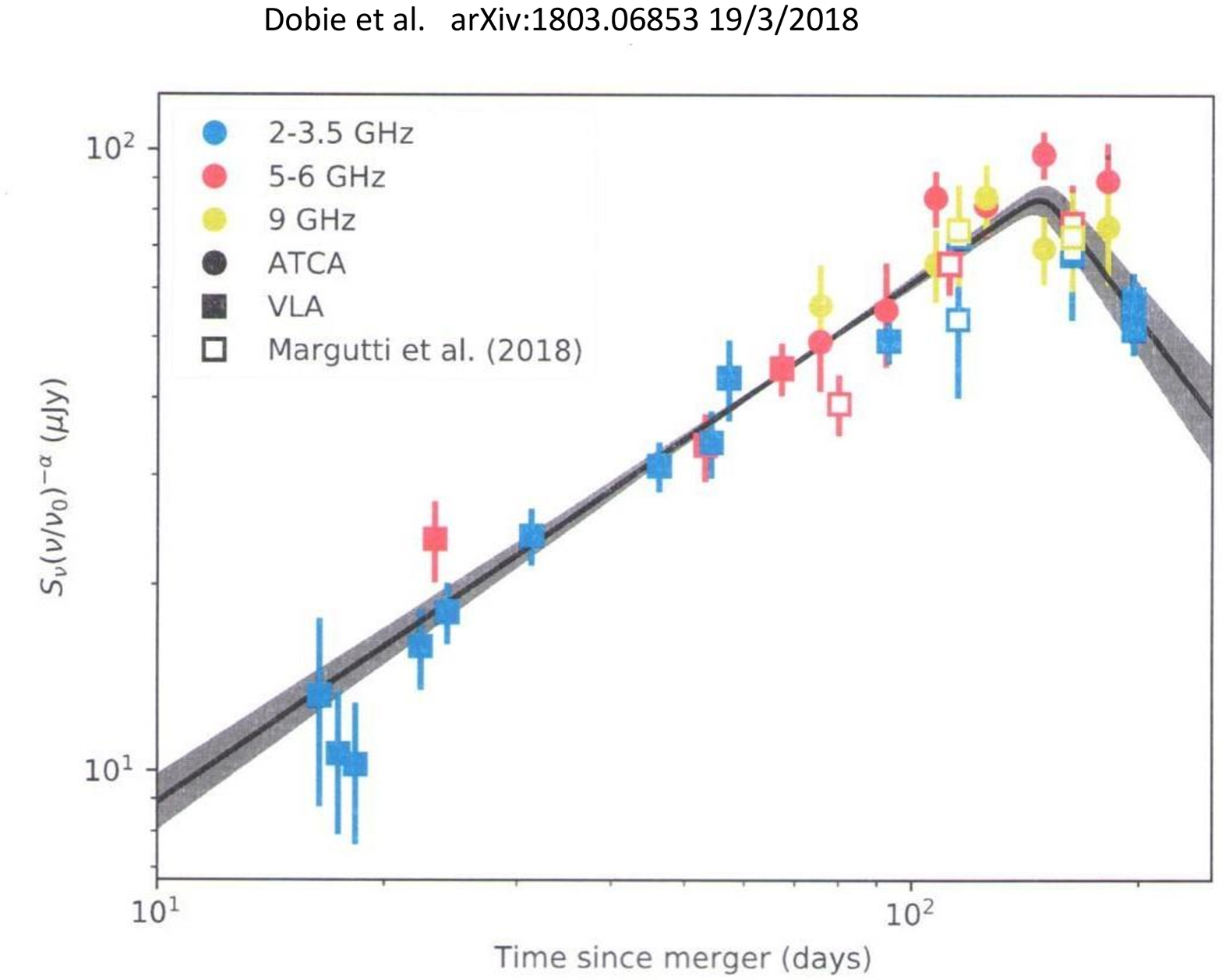,width=10.cm,height=10.cm}
\caption{Figure 1 adapted from [101] show the radio light curve 
of SHB170817A from ATCA (circles) and VLA (squares) observed 
before March 3, 2018, and grouped by frequency band [101].
The flux densities have been adjusted to 5.5 GHz assuming a spectral index
$-0.57\!\pm\!0.04$. Shown also is a  best fit smoothed broken power-law with a
temporal index $0.84\!\pm\!0.05$  on the rise, peak time $149\!\pm\! 2$ d,
and a temporal index $1.6\!\pm\!0.2$ on the decay.} 
\end{figure}
Figure 36 (Figure 2 adapted from [102]) summarizes the 0.6-10 GHz 
observations of the radio afterglow of SHB170817A covering the period 
up to 300 days post-burst, taken with the upgraded Karl G. Jansky Very 
Large Array, the Australia Telescope Compact Array, the Giant Metrewave 
Radio Telescope and the MeerKAT telescope. On the basis of these 
data and its parametrization as a smoothed broken power-law [103] 
with a temporal index $\alpha\!=\!0.84\!\pm\!0.05$ on the rise, 
peak time $149\!\pm\! 2$ day, and a temporal index $1.6\!\pm\!0.2$ 
on the decay, the authors of [102] now concluded that these data consist  
"Strong Jet Signature in the Late-Time Lightcurve of GW170817" in contrast 
to .
\begin{figure}[]
\centering
\epsfig{file=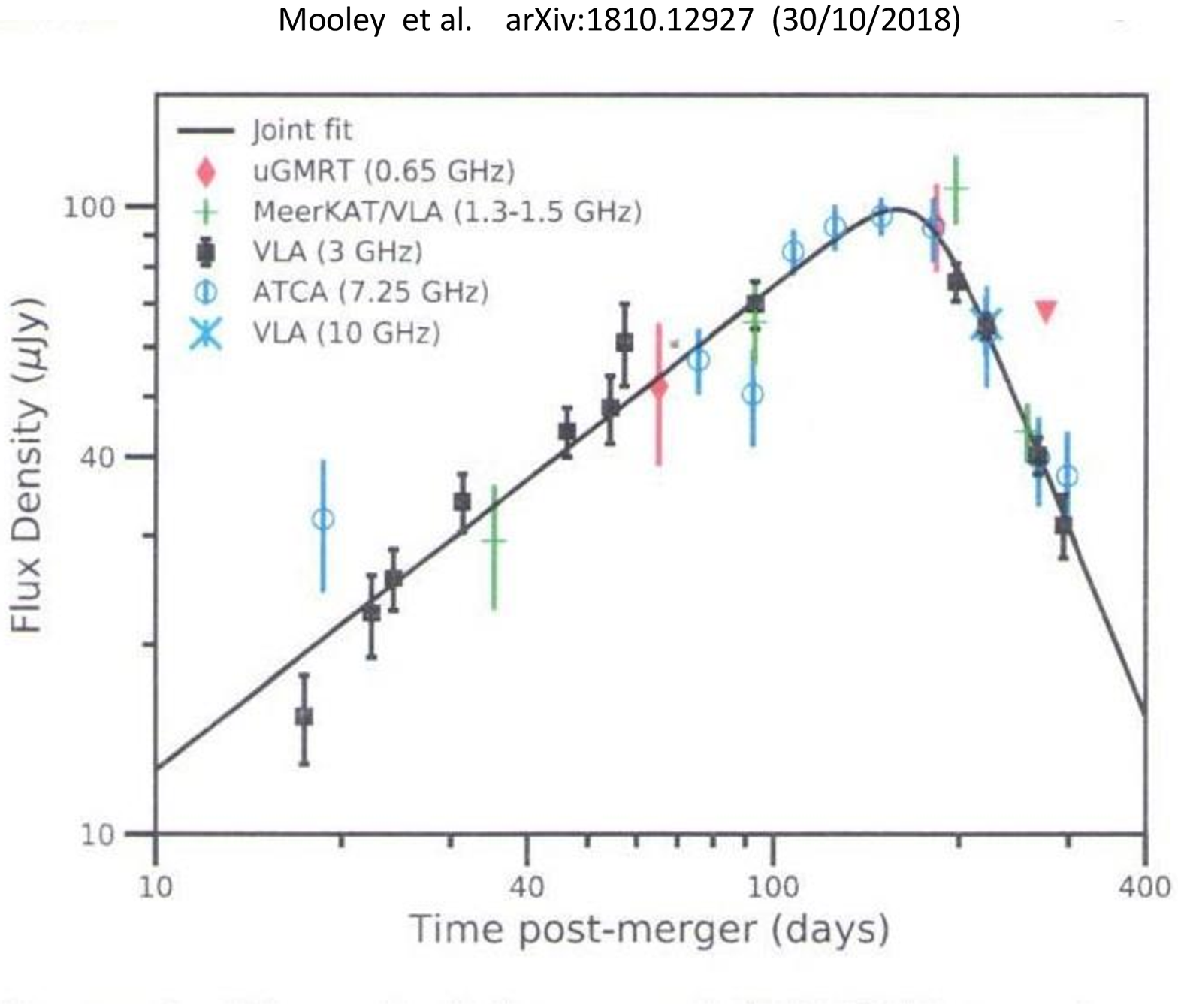,width=10.cm,height=10.cm}
\caption{Figure 2 adapted from [102]. The radio light curve 
of SHB170817A, measured until June 12, 2018,  
spanning multiple frequencies, and scaled to 3 GHz using the 
spectral index $-0.53$. Shown also is 
a best fit smoothly broken power-law parametrization [103]
with five adjustable parameters.}
\end{figure}
The authors justify their conclusion  by the fact 
that a flux density $F_\nu(t)\!\propto\! t^{\alpha}\nu^{\beta}$ with the 
observed $\beta\!=\!0.54$ and the observed values of $\alpha$, both before 
and after the temporal break, cannot be produced by a spherical relativistic 
fireball [102]. On the other hand, for a jet viewed on-axis after the 
jet-break, the power-law decay index is $\alpha\!=\!-\!p$ where 
$\beta\!=\!(1\!-\!p)/2$ 
[104], i.e., $p\!=\!2.17$ for SHB170817A, consistent with that observed. 
However,\\
(a) The last result is valid only for an hypothesized fast spreading jet, i.e. 
a conical jet with a fast lateral expansion ($V_\perp\!\approx\!c$ in the jet 
rest frame at the jet break time [104], which stops its propagation). This
is not supported by the VLBI observations of the radio afterglow of SHB170817A 
[82], which show a compact superluminal source (CB) rather than a spreading 
conical flow with an expanding large opening angle.\\
(b) The relation $\alpha\!=\!-\!p\!=\!1\!-\!2\Gamma$, where $\Gamma$
is the post break photon index, is seldom satisfied in  LGRBs and often yields 
unacceptable $p\!<\! 2$. Due to a large uncertainty,
it is not clear yet whether the late-time afterglow of SHB170817A 
can be  really parametrized well by a broken power-law. 

All types of fireball, firecone, or structured jet models, which have been  
claimed in either posted papers in the arXive, or submitted papers for 
publication in journals, failed to
predict correctly the observed light curves after the  posting or
or submission for publication of these papers.  
This is demonstrated in Figures 37. In Figure 37 Top (Figure 1. adapted from 
V2 of [105]) a structured jet with a relativistic, energetic core surrounded by
slower and less energetic wings  produces an  afterglow emission that brightens
characteristically with time, as was seen in the afterglow of SHB170817A in 
2017, and was postdicted. It was argued that, initially, one  sees only  the 
relatively slow material moving towards us. As time passes, larger and 
larger sections of the outflow become visible, increasing the luminosity of
the afterglow. 
\begin{figure}[]
\centering
\vbox{
\epsfig{file=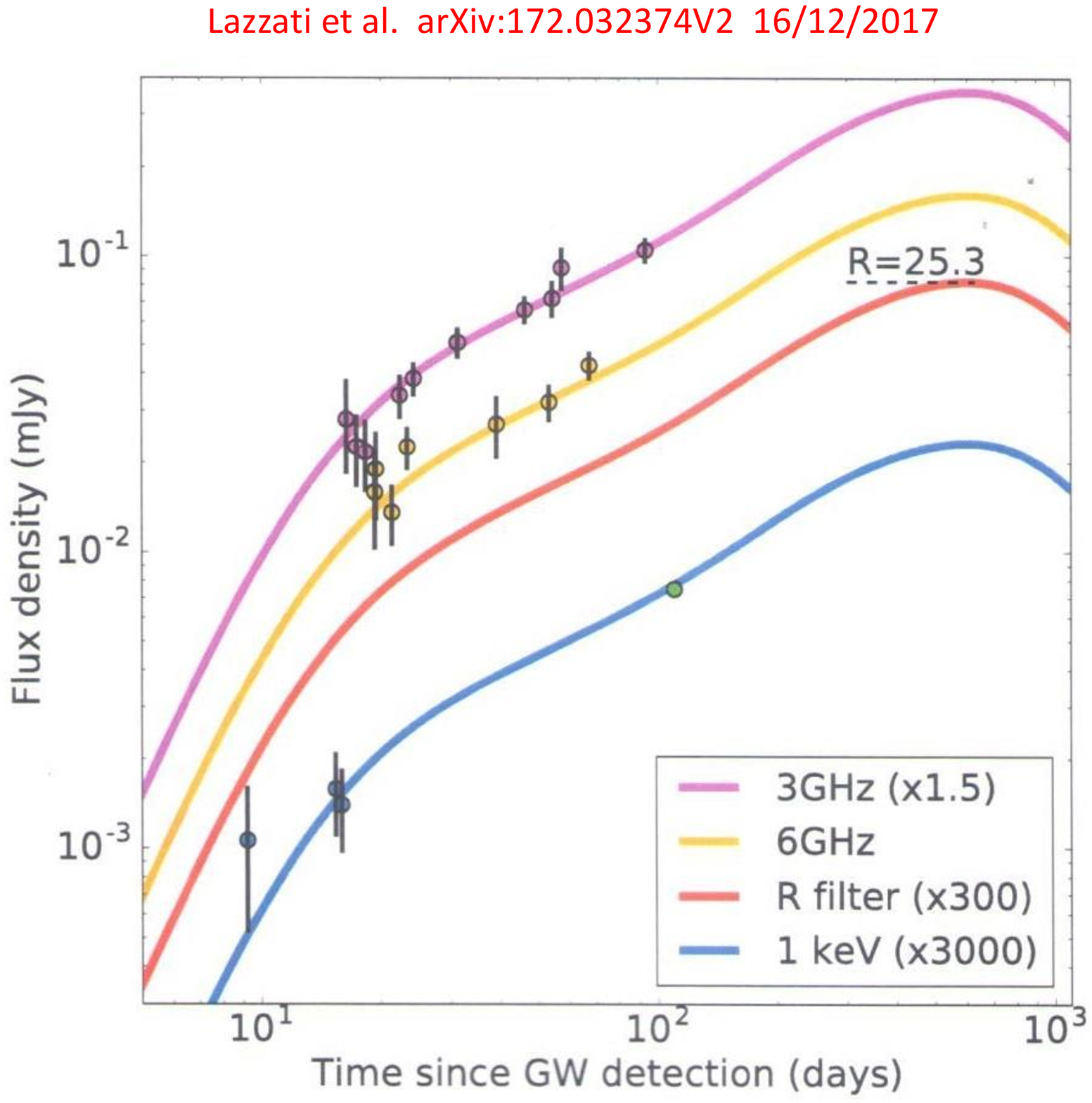,width=10.cm,height=10.cm}
\epsfig{file=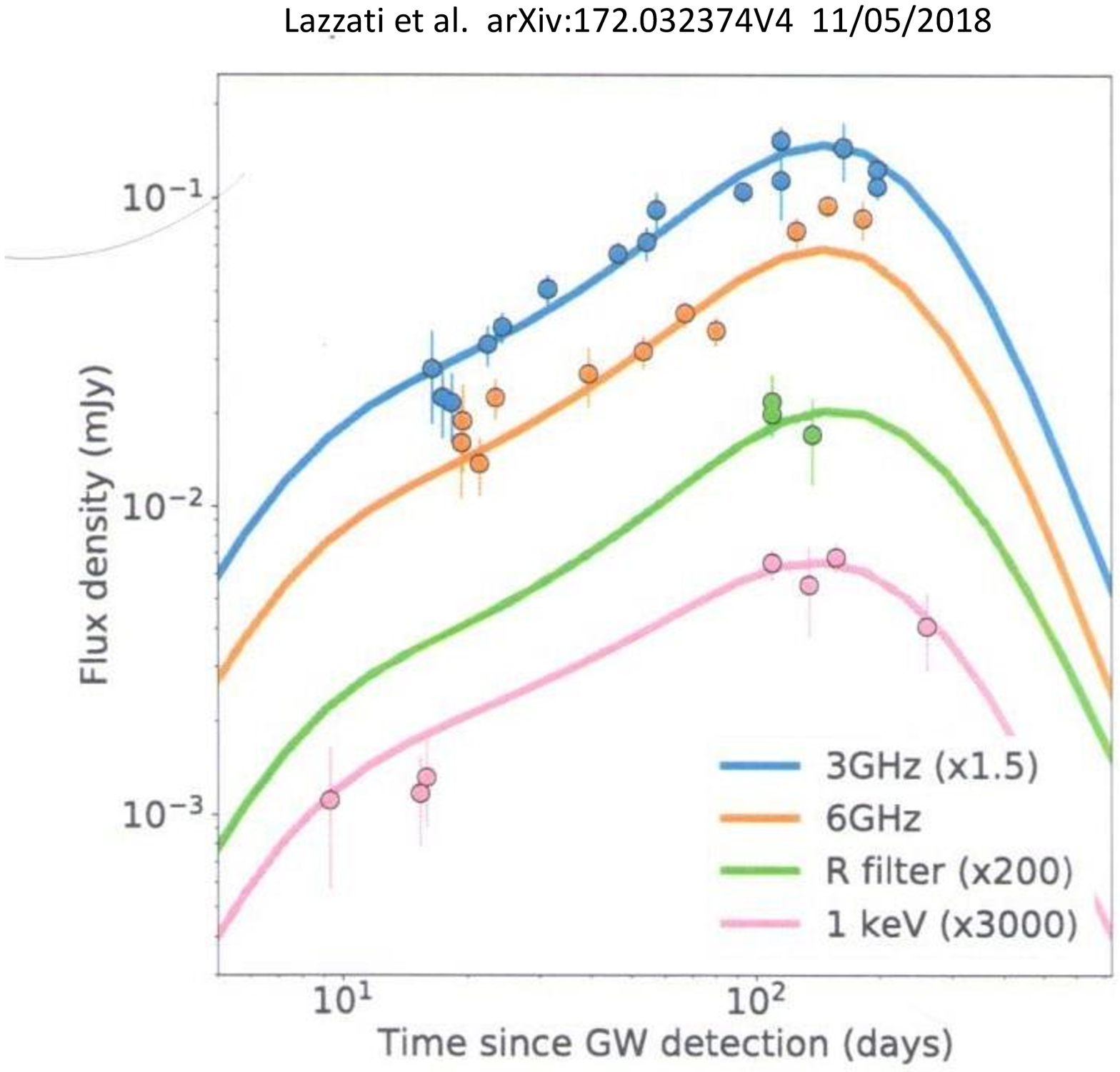,width=10.cm,height=10.cm}
}
\caption{Top: Best fit radio, optical and X-ray (ROX)  light curves of  
an off-axis structured jet model reported in [105] to the ROX 
afterglows of SHB170817A measured before December 2017.
Bottom: Best fit  light curves to
the observed ROX afterglow of SHB170817A until April 2018, obtained from a
structured jet model and reported in [106] (Figure 2 in version 4
of [105]).}
\end{figure}
In the last revised version (V4) of [105], the postdicted light
curves in Figure 37 top  were replaced by best fits to the measured 
radio, optical and X-ray  light
curve until 250 days after burst (March 2018)
shown in Figure 37 bottom (Figure 2 adapted
from [106], the final published version of [105]).\\

\noindent
Despite reproducing light curves with many free adjustable parameters,
conical jets with standard lateral structures have not been shown capable 
of reproducing  cannonball-like VLBI radio
images displaced with a superluminal velocity 
from the location of SHB170817A, as was observed [82] between day 
75 and 230 after burst [82].

\section{Conclusions}
Table III summarizes the confrontations of the key falsifiable predictions 
of the fireball and cannonball models of GRBs, with observations rather 
than with prejudices and beliefs. They clearly 
demonstrate that more than 50 years after the discovery of GRBs, the minority 
views on GRBs continue to be the correct views. This is summarized 
briefly in Table IV. It has been obscured, however, by the continuous flow of 
uncritical papers and reviews of GRBs by biased followers and promoters of the 
fireball model [27]. Moreover, whenever it became clear from observations that 
a minority view is the correct view, it was adapted/incorporated into the FB 
model without much ceremony, proper references, and due credit to its true 
origin!

{}

\newpage
\vspace{3cm}

\begin{table*}
\caption{Critical Tests of The Cannonball and Fireball  models of GRBs and SHBs}
\label{table3}
\centering
\begin{tabular}{l l l l l l}
\hline
\hline
 ~ Test  &~~~~~~~ Cannonball Model~~~~~~~   &    &~~~~~~~~ Fireball Model~~ &             \\
\hline
Test 1   & Large GRB Linear Polarization &{\bf V}~~~~~~& Small GRB  Polarization  & {\bf X } \\
Test 2   & Prompt Emission Correlations  &{\bf V}~~~~~~& Frail Relation       & {\bf X } \\
Test 3   & Univ. Shape of GRB Pulses     &{\bf V}~~~~~~& Curveture Radiation  & {\bf X } \\
Test 4   & SN-GRBs: Canonical Afterglow  &{\bf V}~~~~~~& Canonical AG not expected  &{\bf X}\\
Test 5   & AG break-time correlations    &{\bf V}~~~~~~& AG break-time correlations & {\bf X }\\
Test 6   & Post Break Closure Relation   &{\bf V}~~~~~~& Post break Closure Relation & {\bf X }\\  
Test 7   & Missing Breaks (Too Early)    &{\bf V}~~~~~~& Missing  Breaks (Too Late)  & {\bf X }\\
Test 8   & Predicted Chromatic Behavior  &{\bf V}~~~~~~& Achromatic behavior Predicted & {\bf X }\\
Test 9   & SN-less GRBs:Universal MSP AG &{\bf V}~~~~~~& Magnetar Jet re-energization  & {\bf X }\\
Test 10  & GRB  Rate $\propto$ SFR       &{\bf V}~~~~~~& GRB Rate not  $\propto$ SFR   & {\bf X }\\
Test 11  & LL GRBs = Far Off-axis GRBs   &{\bf V}~~~~~~& LL GRBs = Different GRB class &{\bf X }\\
Test 12  & Super luminal CBs             &{\bf V}~~~~~~& Superluminal Image Expansion&{\bf X}\\
Test 13  & Properties of SHB170817A      &{\bf V}~~~~~~& Properties of SHB170817A       &{\bf X}\\
\hline
\end{tabular}
\end{table*}  
\vspace{3cm}

\newpage
\begin{table*}
\caption{Majority and minority views on GRBs before decisive observational evidence}
\label{table4}
\centering
\begin{tabular}{l l l l l l}
\hline  
\hline
~~Key property      & ~~~~~~~Majority  View&~~~~~ &~~~~~~~~~~Minority View&~~~ \\ 
\hline 
GRB Location        & Galactic  &X~~~~~ &Extragalactic& V\\ 
GRBs Produced By    & Relativistic Fireballs& X~~~~~  & Highly Relativistic Jets &V  \\
GRB Pulses from     & Collisions of $e^+e^-$ shells &X~~~~~&ICS of External Light by plasmoids&V\\ 
Emission Geometry   & Isotropic  & X~~~~~              & Narrowly Beamed & V \\
\hline
Afterglow Origin    & Shocked ISM & X~~~~~ & Synchrotron From ISM Swept Into Jet &V \\ 
Afterglow Distribution  & Isotropic &X~~~~~  & Narrowly Beamed &V\\
\hline
LGRBs Origin        & SN-Less Collapse to BH&X~~~~~ & Stripped Envelope SN &V\\    
SN1998bw/GRB980425  & Rare SN/Rare GRB &X~~~~~ & SNIc-GRB Viewed Far Off-Axis &V \\   
LL GRBs             & Different class of GRBs &X~~~~~ &  Ordinary GRBs Viewed Far Off-Axis&V\\ 
SN-Less LGRBs       & SN-Less Collapse to BH   & X~~~~~ & n* to q* phase transition in HMXBs&V \\
\hline
Prompt Emission     & Synchrotron &X~~~~~ & Inverse Compton &V \\    
Origin of Jet break & Conical Jet Deceleration &X~~~~~& Plasmoid Deceleration by Swept-in ISM&V\\  
Rate of GRBs        & $\propto$ SFR + Evolution &X~~~~~ & $\propto$ SFR + beaming &V\\  
AG plateau (SN-GRBs)    & Jet Re-energization &X~~~~~ & Jet Deceleration at 
Early Time &V\\ 
AG plateau(SN-less GRBs)& Jet Re-energization &X~~~~~ & PWN Emission Powered 
by MSP &V\\
Missing Jet Breaks  & After AG Observations End  &X~~~~~& Before AG Is Observable \\ 
\hline
\end{tabular}
\end{table*}

\end{document}